\documentclass[apjl]{emulateapj}
\pdfoutput=1

\usepackage{graphicx}
\usepackage{t1enc}
\usepackage[varg]{txfonts}
\usepackage{epsfig}
\usepackage{rotating}
\usepackage{natbib}
\usepackage{color}

\newcommand{\cmt}   {cm$^{-3}$}
\newcommand{\jpb}   {$\rm Jy~beam^{-1}$}    

\newcommand{\mo}    {$M_{\sun}$}

\newcommand{\nh}    {NH$_3$}

\newcommand{\eg}    {e.\,g.,}
\newcommand{\ie}    {i.\,e.,}

\definecolor{RED}{rgb}{1.0,0.0,0.0}

\shorttitle{ALMA and VLA observations of OMC-1S}
\shortauthors{Palau et al.}

\begin{document}

\title{Thermal Jeans fragmentation within $\sim1000$ AU in OMC-1S}

\author{Aina Palau\altaffilmark{1}, 
Luis A. Zapata\altaffilmark{1},
Carlos G. Rom\'an-Z\'u\~niga\altaffilmark{2},
\'Alvaro S\'anchez-Monge\altaffilmark{3},
Robert Estalella\altaffilmark{4},
Gemma Busquet\altaffilmark{5},
Josep M. Girart\altaffilmark{5},
Asunci\'on Fuente\altaffilmark{6},
Benoit Commer\c con\altaffilmark{7}
}
\altaffiltext{1}{Instituto de Radioastronom\'ia y Astrof\'isica, Universidad Nacional Aut\'onoma de M\'exico, P.O. Box 3-72, 58090, Morelia, Michoac\'an, M\'exico}\email{a.palau@irya.unam.mx}
\altaffiltext{2}{Instituto de Astronom\'ia, Universidad Nacional Aut\'onoma de M\'exico, Apartado Postal 106, 22800 Ensenada, Baja California, M\'exico}
\altaffiltext{3}{I. Physikalisches Institut der Universit\"at zu K\"oln, Z\"ulpicher Strasse 77, 50937 K\"oln, Germany}
\altaffiltext{4}{Departament de F\'isica Qu\`antica i Astrof\'isica, Institut de Ci\`encies del Cosmos (ICC), Universitat de Barcelona (IEEC-UB), Mart\'i Franqu\`es 1, 08028 Barcelona, Catalonia, Spain}
\altaffiltext{5}{Institut de Ci\`encies de l'Espai (IEEC-CSIC), Campus UAB, Carrer de Can Magrans s/n, 08193 Cerdanyola del Vall\`es, Catalonia, Spain}
\altaffiltext{6}{Observatorio Astron\'omico Nacional, Apartado Postal 112, 28803 Alcal\'a de Henares, Madrid, Spain}
\altaffiltext{7}{\'Ecole Normale Sup\'erieure de Lyon, CRAL, UMR CNRS 5574, Universit\'e Lyon I, 46 All\'ee d'Italie, 69364 Lyon Cedex 07, France}

\begin{abstract}
We present subarcsecond 1.3 mm continuum ALMA observations towards the Orion Molecular Cloud 1 South (OMC-1S) region, down to a spatial resolution of 74~AU, which reveal a total of 31 continuum sources.
We also present subarcsecond 7~mm continuum VLA observations of the same region, which allow to further study fragmentation down to a spatial resolution of 40~AU. 
By applying a Mean Surface Density of Companions method we find a characteristic spatial scale at $\sim560$~AU, and we use this spatial scale to define the boundary of 19 `cores' in OMC-1S as groupings of millimeter sources. We find an additional characteristic spatial scale at $\sim2900$~AU, which is the typical scale of the filaments in OMC-1S, suggesting a two-level fragmentation process.
We measured the fragmentation level within each core and find a higher fragmentation towards the southern filament. In addition, the cores of the southern filament are also the densest (within 1100~AU) cores in OMC-1S. This is fully consistent with previous studies of fragmentation at spatial scales one order of magnitude larger, and suggests that fragmentation down to 40~AU seems to be governed by thermal Jeans processes in OMC-1S.
\end{abstract}


\section{Introduction \label{si}}

The processes regulating the fragmentation of molecular clouds are still a matter of vigorous discussion and debate. These processes have been studied from an observational point of view at different spatial scales. At large-scales ($\sim1$~pc) or clump scales, gravoturbulent fragmentation (or thermal Jeans fragmentation) seems to describe well the observations (\eg\ Rom{\'a}n-Z{\'u}{\~n}iga et al. 2009,Takahashi et al. 2013; Walker-Smith et al. 2013; Samal et al. 2015; Kainulainen et al. 2016; Sharma et al. 2016; Hacar et al. 2017b), although there are cases where the effects of feedback (Busquet et al. 2016), or large-scale collapse (Csengeri et al. 2017) seem to be important as well. The term of `gravoturbulent fragmentation' refers to the fragmentation that takes place in a turbulent and self-gravitating cloud. In these clouds, turbulence generates high-density regions where the Jeans mass decreases and thus fragmentation is favored (\eg\ Padoan \& Nordlund 2002; Schmeja et al. 2004; Hopkins 2013; Gong \& Ostriker 2015).

At medium-scales ($\sim0.1$~pc), fragmentation also seems to be dominated by gravity acting in a turbulent medium (\eg\ Teixeira et al. 2016, Kirk et al. 2017). In this respect, Palau et al. (2014, 2015) compile a sample of 19 massive dense cores and study their fragmentation level against different parameters, finding that the fragmentation level seems to be well correlated with the density (calculated within the approximate core size of 0.1~pc), a result consistent with thermal Jeans fragmentation. However, there are also works showing that at these scales the mass of the fragments seem to be too high compared to the thermal Jeans mass (\eg\  Zhang et al. 2009, 2015; Bontemps et al. 2010; Wang et al. 2011, 2014; Pillai et al. 2011; Lu et al. 2015). This needs to be further investigated with better mass sensitivities (Henshaw et al. 2017) and in relation to the kinematics of the surrounding molecular gas (\eg\ Duarte-Cabral et al. 2013).

At even smaller-scales ($\sim0.01$~pc or $\sim1000$~AU), the `fragmentation properties' are usually referred to as `multiplicity properties', and most studies have been carried out towards low-mass star-forming regions. At these scales, new ingredients such as rotational fragmentation (\eg\ Chen et al. 2012; Chabrier et al. 2014) or fragmentation of a (pseudo)disk through gravitational instability (\eg\ Kratter \& Matzner 2006; Tobin et al. 2016) have been proposed to explain the formation of multiple/binary systems.
Thus, it is not clear whether the dominant process regulating fragmentation at these small scales is inherited from large scales (as suggested by some observational works, Lee et al. 2015), or the aforementioned processes start to be important.
At these scales, there is a lack of observations about the fragmentation properties in intermediate and high-mass star-forming regions at very early stages, but is quite well stablished that the multiplicity fraction in high-mass main-sequence stars, as well as the number of companions per system, increases with stellar mass (\eg\  Chini et al. 2012, Kraus et al. 2017). Thus, studying the fragmentation/multiplicity properties in deeply embedded intermediate/high-mass star forming regions should provide clues for the dominant processes determining fragmentation at these scales.

\begin{figure*}
\begin{center}
\begin{tabular}[b]{c}
    \epsfig{file=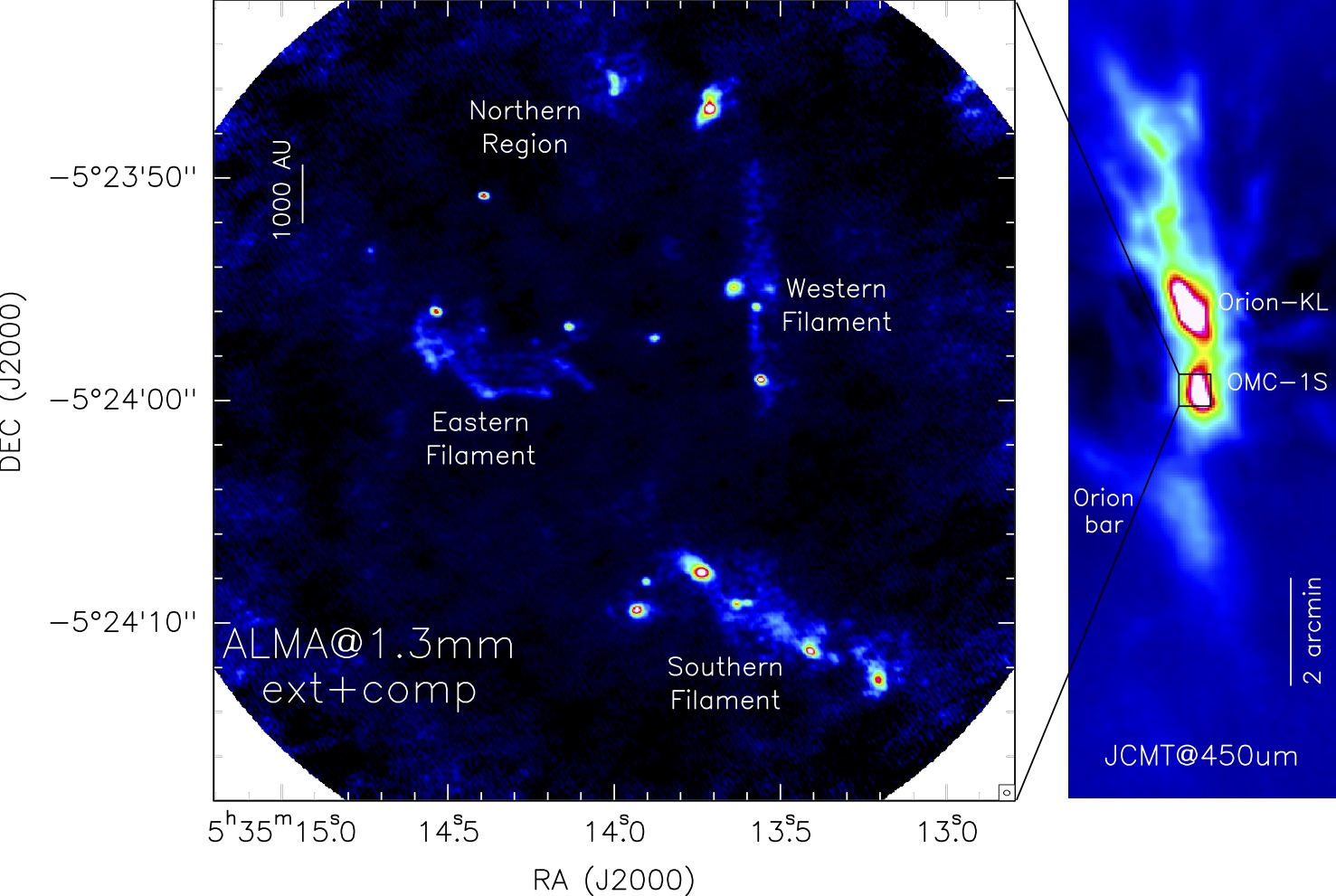, scale=0.35,angle=0}\\
\end{tabular}
\caption{
\emph{Left:} ALMA 1.3~mm continuum emission for the OMC-1S region for extended+compact configurations. The ALMA beam, of $0.28''\times0.22''$, is shown in the bottom-right corner; a scale corresponding to 1000~AU is shown in the top-left corner.
\emph{Rigth:} James Clerk Maxwell Telescope (JCMT) at 450~$\mu$m from di Francesco et al. (2008) showing the large field of view where OMC-1S is embedded, in relation to Orion-KL and the Orion bar. The large-scale filament extending from north to south corresponds to the central part of the Integral Shaped Filament. The black square indicates de field of view of the ALMA data shown on the left panel.
}
\label{falmaall}
\end{center}
\end{figure*}

Here we present a study of the fragmentation properties of a region forming intermediate and low-mass protostars, the Orion Molecular Cloud 1 South (hereafter, OMC-1S). 
The region lies near the center of the Integral Shaped Filament (\eg\ Johnstone \& Bally 1999, Stutz \& Gould 2016), a $\sim50'$ long filament in the Orion A cloud whose central part harbors the famous Orion-KL and Orion bar regions. The right panel of Figure~\ref{falmaall} presents the central $15'$ of the Integral Shaped Filament, with OMC-1S highlighted with respect to Orion-KL and the Orion bar.
%
The most recent measurement of the distance has been done by Kounkel et al. (2017), who report 388~pc. OMC-1S has been widely studied using multiple observational techniques, which have revealed a nascent cluster characterized by a large number of spectacular molecular outflows and HH objects (\eg\ Smith et al. 2004, Zapata et al. 2005, 2006; Henney et al. 2007). 
In addition, the outer layers of the OMC-1S region, which are bathed by external irradiation from the surrounding massive stars within the Orion Nebular Cluster (Tahani, Plume \& Bergin 2016), also seem to be undergoing global infall (\eg\ Hacar et al. 2017a). Therefore, a very active phase of clustered star formation is taking place in OMC-1S, and the region becomes an excellent target to study fragmentation down to very small scales ($\lesssim1000$~AU). In Section 2, we present the millimeter ALMA and VLA observations towards OMC-1S down to spatial scales of $\sim40$~AU for the VLA observations. In Section 3, we describe the main results. In Section 4, we analyse the properties of the fragmenting cores in OMC-1S; in Section 5 we discuss our main findings, and in Section 6 we give the conclusions. 

\begin{figure*}
\begin{center}
\begin{tabular}[b]{c}
    \epsfig{file=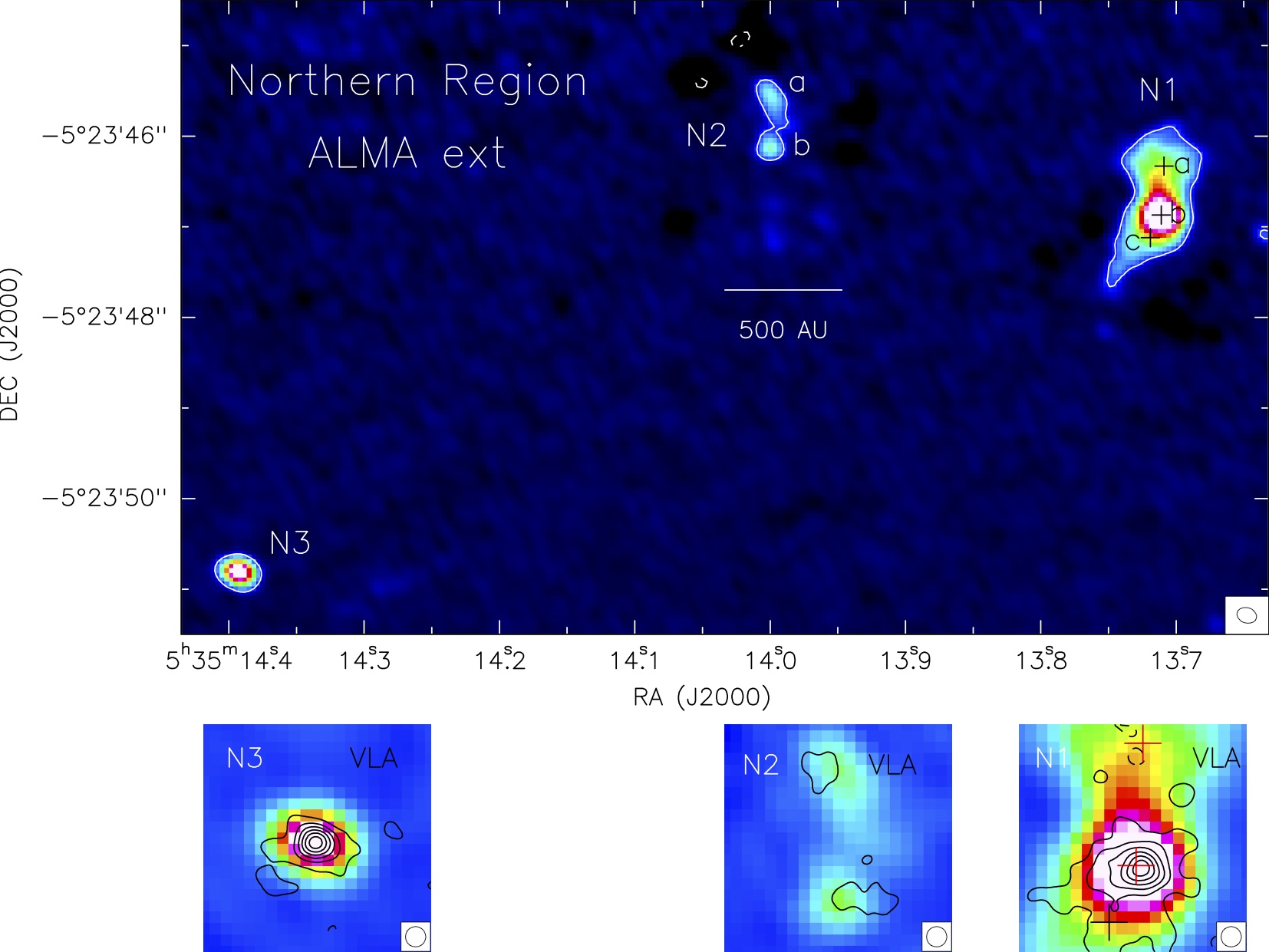, scale=0.3,angle=0}\\
\end{tabular}
\caption{Top panel: ALMA 1.3~mm continuum emission towards the north of the OMC-1S region, using only the extended configuration. White (dashed) contours correspond to the 12$\sigma$ ($-12\sigma$) emission. The ALMA beam is shown in the bottom-right corner. Lower panels: colorscale: idem as top panel; black contours: VLA 7~mm emission in a zoom of $1''\times1''$ centered on each source identified in the ALMA image. Contours are $-2.5$ (dashed), 2.5, 5, 7, 9, 11, 13, 15, and 17 times the rms of the image, 0.13~m\jpb. Note that the 7~mm detection threshold is taken at 5$\sigma$, corresponding to two contours in the figures.
In all the lower panels, the VLA beam is shown in the bottom-right corner. Plus signs in the top panel and in the bottom-right panel correspond to the peaks of the 3 Gaussians required to model the ALMA source N1.}
\label{fvlan}
\end{center}
\end{figure*}

\section{Observations \label{sobs}}

\subsection{ALMA at 1.3~mm}

The ALMA\footnote{ALMA is a partnership of ESO (representing its member states), 
NSF (USA) and NINS (Japan), together with NRC (Canada) and NSC and ASIAA (Taiwan) 
and KASI (Republic of Korea), in cooperation with the Republic of Chile. 
The Joint ALMA Observatory is operated by ESO, AUI/NRAO and NAOJ.}  Band 6 (1.3 mm) observations were 
carried out on 2015 May 19 and August 18 as part of the Cycle 2 program 2013.1.01037.S. 
On May 19, we used 38 of the 12 m antennas, yielding baselines with projected lengths from 21 to 555~m (16--427~k$\lambda$), while 
on August 18 we used 37 antennas also of 12 m, yielding baselines with projected lengths from 43 to 1600~m (33--1230~k$\lambda$).  
Including both epochs, the covered range of projected baselines is 16--1230~k$\lambda$, which corresponds to a Largest Angular Scale to which ALMA was sensitive of $5.7''$ (following equation A5 of Palau et al. 2010). 
The integration time on-source (OMC-1S) was about 8 minutes in each observation.  

The phase center for both observations was the same $\alpha(J2000) = 05^h~ 35^m~ 14\rlap.^s0$;
$\delta(J2000) = -05^\circ~ 24'~ 00''$. The continuum image was obtained averaging line-free channels
from four spectral windows (of 1.875 GHz width) centered at rest frequencies: 233.683 GHz,  231.323 
GHz, 219.266 GHz, and 216.875 GHz, which covers a total bandwidth of 7.5 GHz. 

The weather conditions were very good and stable with an average precipitable water vapor of 0.8 mm 
and an average system temperature of 60 K. The ALMA calibration included simultaneous observations of the 
183 GHz water line with water vapor radiometers, used to reduce the atmospheric phase noise. 
Quasars J0541$-$0541, J0522$-$364, and J0750$+$1231 were used to calibrate the bandpass, the flux scale, the atmosphere 
and the gain fluctuations. 

The data were calibrated, imaged, and analyzed using the Common Astronomy Software Applications (CASA) version 4.7. 
Imaging of the calibrated visibilities was done using the task CLEAN. We combined the two epochs with the task 
CONCAT in CASA. The resulting (dynamic-range limited) image rms noise was 0.7 mJy beam$^{-1}$ at an angular resolution of $0\rlap.{''}28 \times 0\rlap.{''}22$
with PA = 81.64$^\circ$. 
We also generated an image using only the extended configuration, in order to extract the compact sources. In this case, the rms of the 1.3~mm continuum is 0.45~m\jpb, and the synthesized beam is  $0\rlap.{''}22 \times 0\rlap.{''}17$, corresponding to a spatial resolution of $\sim74$~AU at the distance of OMC-1S, with PA=73.37$^\circ$. Throughout the paper, we will refer to the rms noise of the extended configuration image by $\sigma$. 
We used the ROBUST parameter of CLEAN in CASA set to 0
for an optimal compromise between angular resolution and sensitivity. The millimeter continuum image obtained 
for OMC-1S was corrected by the primary beam attenuation. The primary beam at this wavelength had a full width at half maximum of about $27^{\prime\prime}$.  


\begin{figure*}
\begin{center}
\begin{tabular}[b]{c}
    \epsfig{file=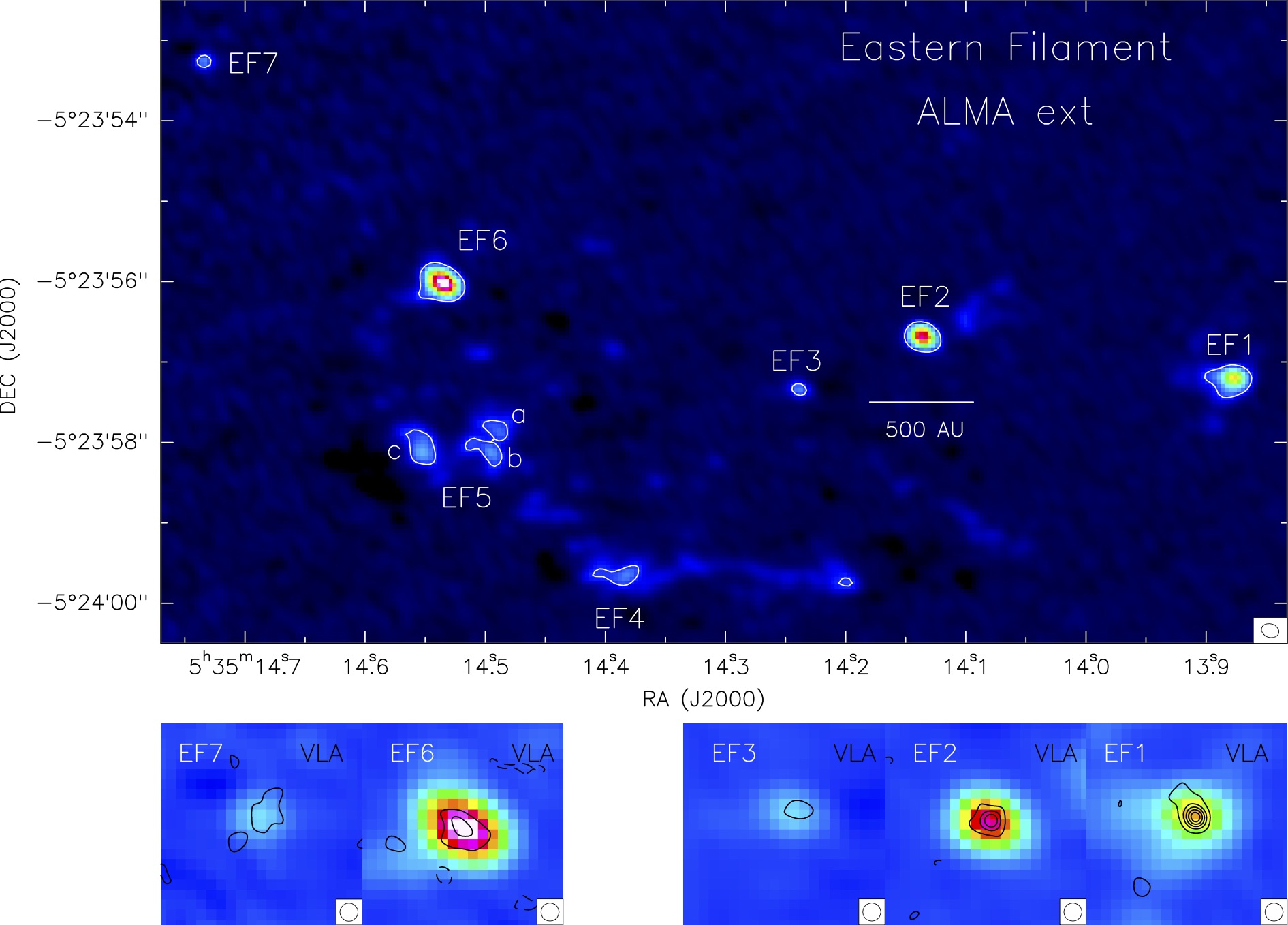, scale=0.26,angle=0}\\
\end{tabular}
\caption{Same as Fig.~\ref{fvlan} for the Eastern Filament of OMC-1S.
}
\label{fvlaef}
\end{center}
\end{figure*} 

\begin{figure}
\begin{center}
\begin{tabular}[b]{c}
    \epsfig{file=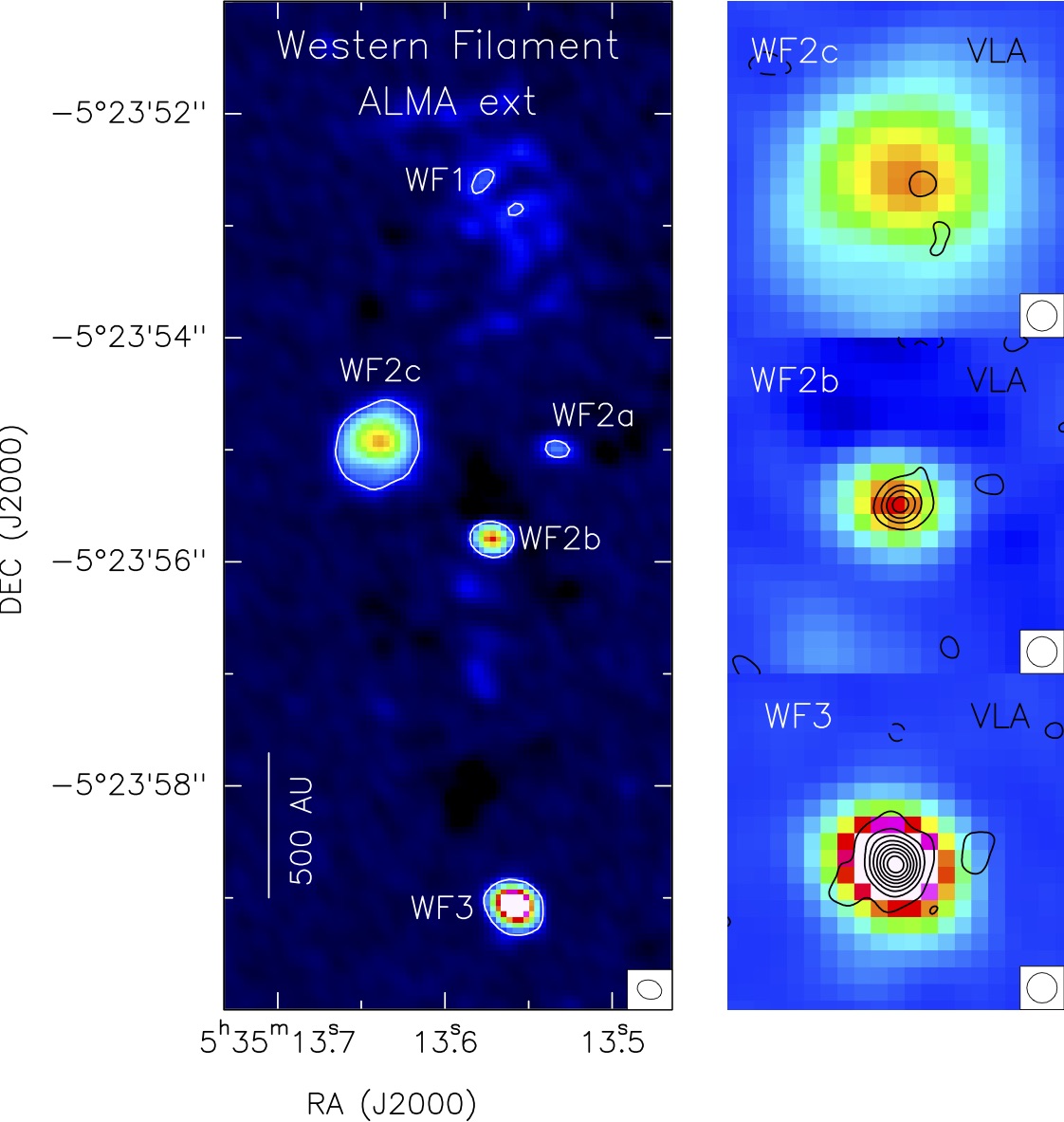, scale=0.21,angle=0}\\
\end{tabular}
\caption{Same as Fig.~\ref{fvlan} for the Western Filament of OMC-1S.
}
\label{fvlawf}
\end{center}
\end{figure} 

\begin{figure*}
\begin{center}
\begin{tabular}[b]{c}
    \epsfig{file=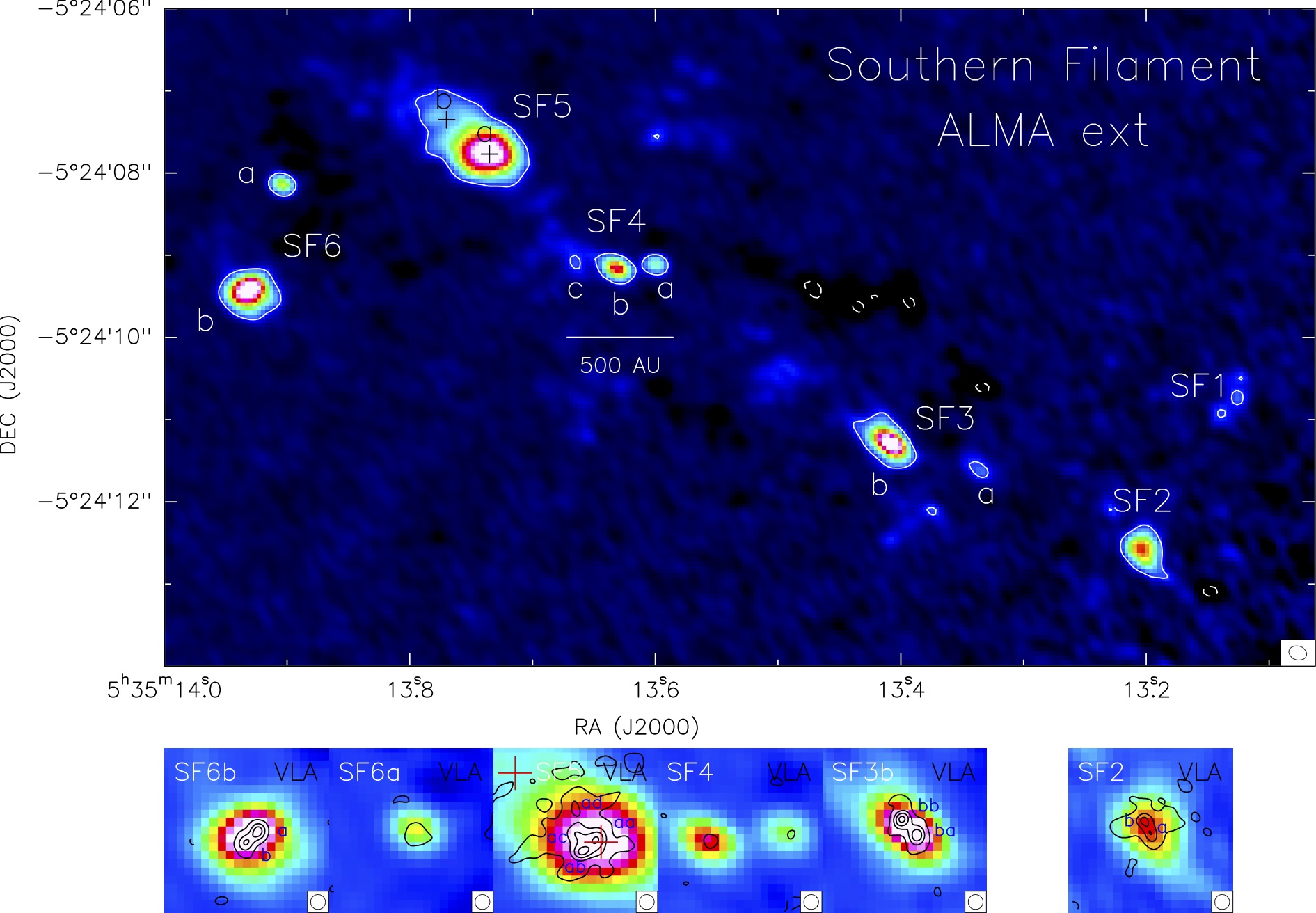, scale=0.27, angle=0}\\
\end{tabular}
\caption{Same as Fig.~\ref{fvlan} for the Southern Filament of OMC-1S.
For the case of SF2, and SF5 we additionally show the contour at 6$\sigma$ to make the structure clearer.  
Plus signs in the top panel and in the third lower panel correspond to the peaks of the 2 Gaussians required to model the ALMA source SF5.
}
\label{fvlasf}
\end{center}
\end{figure*}

\subsection{VLA at 7~mm}

The observations were made with the Very Large Array of NRAO\footnote{The National 
Radio Astronomy Observatory is a facility of the National Science Foundation operated
under cooperative agreement by Associated Universities, Inc.} 
in the continuum mode at 7 mm (43.34 GHz) during 2004 November 10, under the project code AZ154. 
The effective bandwidth of the observations was 100 MHz. 
At this epoch, the VLA was in A configuration, providing baselines covering projected lengths from 0.68 to 36 km (99--5200~k$\lambda$). 
This corresponds to a Largest Angular Scale to which the VLA was sensitive of $0.9''$.

The phase center of the observations was $\alpha(J2000) = 05^h~ 35^m~ 14\rlap.^s0$;
$\delta(J2000) = -05^\circ~ 24'~ 03''$, very close to the center of the radio cluster reported by Zapata et al. (2004). 
The absolute amplitude calibrator was J1331$+$ 305 (with an adopted flux density of 1.45 Jy) and the 
phase calibrator was J0541$-$056 (with a bootstrapped flux density of 1.10 $\pm$ 0.09).

The data were acquired and reduced using the recommended VLA  procedures for high-frequency  data,  
including  the  fast-switching mode with a cycle of 120 s. We used the ROBUST parameter of IMAGR in AIPS set to 0. 
The resulting image had an angular resolution of $0\rlap.{''}065 \times 0\rlap.{''}053$ with PA = $-7.73^\circ$, but we smoothed the image to a final circular beam of $0.09''$ in order to be more sensitive to faint and extended structure.  The rms noise of the smoothed image is 0.13~m\jpb.
The 7 mm continuum image was also corrected by the primary beam attenuation. The primary beam at this wavelength has 
a full width at half maximum of about $60^{\prime\prime}$. 


\section{Results \label{sres}}

In Fig.~\ref{falmaall} we present the ALMA 1.3~mm continuum emission towards OMC-1S with an angular resolution of $\sim0.25''$, corresponding to $\sim100$~AU. 
The strongest millimeter sources detected by ALMA coincide very well with the sources already detected with the Submillimeter Array (see, \eg\ Figure 1 of Zapata et al. 2005), but the ALMA image, about two orders of magnitude deeper, reveals additionally a number of faint point-like sources as well as a striking extended emission around the strong millimeter sources. We identify three main filamentary structures, plus three additional sources to the north of the image, which do not seem to belong to any filamentary structure. Two of the main filaments lie in the central part of the image, with the eastern one (hereafter, the `Eastern Filament') elongated in the east-west direction, and the western one (the `Western Filament') elongated in the north-south direction. The third filament, which is the most prominent one, lies about 10$''$ to the south (hereafter, the `Southern Filament') and is also elongated in the east-west direction.

In order to extract the point and fainter sources detected by ALMA, we generated an image using only the extended configuration data. In Figs.~\ref{fvlan}, \ref{fvlaef}, \ref{fvlawf} and \ref{fvlasf} we show a zoom of the ALMA image of Fig.~\ref{falmaall} on the four different regions/filaments identified in this figure. We identified 31 compact sources above a 12$\sigma$  threshold\footnote{The ALMA array is so sensitive that it is detecting many extended emission in OMC-1S. Because of this, some secondary lobes of the dirty beam cannot be properly removed with the cleaning process. Those secondary lobes reach in some cases the 12$\sigma$ contour, as is the case of the Southern Filament and the northern part of N2. 12$\sigma$ is the threshold where the secondary lobes are reduced to sizes smaller than the beam size.} (white contour in the figures; in this identification, we did not take into account those sources whose 12$\sigma$ contour encompasses an area smaller than half of the beam so that we avoid artifacts or peaks of noise). For the case of N1 and SF5, which are the strongest and most extended millimeter sources in the region, we identified subcomponents (using only the extended configuration data) by fitting 2D Gaussians and inspecting the residuals: if the residuals lie above the 12$\sigma$ detection threshold, we required the inclusion of an additional Gaussian. 
Among the 31 identified sources, 11 sources belong to the Southern Filament (named SF1 to SF6, with SF3, SF4, SF5 and SF6 consisting of several components); 9 sources belong to the Eastern Filament (named EF1 to EF7, with EF5 consisting of 3 components), 6 sources belong to the northern region (N1 to N3, with N1 consisting of 3 components and N2 consisting of 2 components), and 5 sources belong to the Western Filament (named WF1 to WF3, with WF2 consisting of 3 components). 
The Eastern Filament seems to have two `arms', the northern arm including EF1, EF2, and EF3, and the southern arm including EF4 and EF5, with EF6 and EF7 being common sources to both arms. Since EF4 and EF5 are interconnected by extended emission which reaches the EF6 and EF2 sources (see Figs.~\ref{falmaall} and \ref{fvlaef}), it seems reasonable to think that the two arms should have a common origin. 

\begin{table*}
\caption{Parameters of the sources detected with ALMA at 1.3~mm in OMC-1S}
\begin{center}
{\small
\begin{tabular}{lcccccccc}
\noalign{\smallskip}
\hline\noalign{\smallskip}
&\multicolumn{2}{c}{Position$^\mathrm{b}$}
&$I_\mathrm{\nu}^\mathrm{1.3mm}$~$^\mathrm{b}$
&$S_\mathrm{\nu}^\mathrm{1.3mm}$~$^\mathrm{b}$
&Dec. size~$^\mathrm{b}$
&Dec. PA~$^\mathrm{b}$
&Mass$^\mathrm{c}$
\\
\cline{2-3}
Source
&$\alpha (\rm J2000)$
&$\delta (\rm J2000)$
&(m\jpb)
&(mJy)
&($''\times''$)
&(\degr)
&(\mo)
&Association$^\mathrm{d}$\\
\noalign{\smallskip}
\hline\noalign{\smallskip}
N1a         	&05:35:13.709   &$-05.23.46.33$ 	&$44\pm2$	&$293\pm19$	&$0.65\times0.51$	&128		&0.76			&-\\
N1b         	&05:35:13.711   &$-05.23.46.87$ 	&$118\pm3$	&$279\pm10$	&$0.30\times0.26$	&7		&0.73			&137--347\\
N1c         	&05:35:13.719   &$-05.23.47.12$ 	&$45\pm2$	&$419\pm22$	&$0.96\times0.50$	&129		&1.09			&-\\
N2a       	&05:35:13.999   &$-05.23.45.63$ 	&$30\pm2$	&$102\pm9$	&$0.62\times0.16$	&20 	    	&0.27			&-\\
N2b       	&05:35:13.998   &$-05.23.46.08$ 	&$33\pm2$	&$97\pm8$ 	&$0.40\times0.27$	&154    	&0.25			&-\\
N3   		&05:35:14.392   &$-05.23.50.82$	&$93\pm1$   	&$114\pm2$	&$0.13\times0.10$	&94 	     	&0.30			&144--351\\
\hline
EF1         	&05:35:13.878   &$-05.23.57.21$ 	&$35\pm1$   	&$62\pm2$	&$0.27\times0.15$	&109 	&0.16			&139--357\\ 
EF2         	&05:35:14.135   &$-05.23.56.69$	&$57\pm1$   	&$76\pm2$  	&$0.15\times0.13$	&84		&0.20			&141--357\\ 
EF3		&05:35:14.238	 &$-05.23.57.38$	&$8.4\pm0.3$	&$13.8\pm0.8$	&$0.22\times0.17$	&102		&0.04			&-\\
EF4		&05:35:14.386	 &$-05.23.59.64$	&$16\pm1$	&$197\pm16$	&$1.34\times0.49$	&61		&0.51			&-\\
EF5a	&05:35:14.497	 &$-05.23.57.84$	&$10\pm1$	&$48\pm8$	&$0.56\times0.40$	&50		&0.12			&-\\
EF5b	&05:35:14.510	 &$-05.23.58.02$	&$9\pm2$		&$24\pm7$	&$0.52\times0.18$	&91		&0.06			&-\\
EF5c		&05:35:14.552	 &$-05.23.58.06$	&$18\pm1$	&$176\pm12$	&$1.10\times0.45$	&31		&0.46			&-\\
EF6         	&05:35:14.535   &$-05.23.56.02$	&$84\pm2$     	&$134\pm5$ 	&$0.22\times0.16$	&53		&0.35			&145--356\\ 
EF7		&05:35:14.734	 &$-05.23.53.28$	&$14.6\pm0.6$	&$15\pm1$ 	&$-\times-$		&$-$		&0.04			&\\
\hline
WF1		&05:35:13.578	 &$-05.23.52.50$	&$4.0\pm0.6$	&$19\pm4$ 	&$0.73\times0.25$	&144		&0.05			&-\\
WF2a    	&05:35:13.531   &$-05.23.55.00$ 	&$14.1\pm0.8$	&$25\pm2$ 	&$0.26\times0.17$	&68 		&0.06			&-\\ 
WF2b     	&05:35:13.571   &$-05.23.55.81$ 	&$50\pm1$      &$65\pm3$	&$0.14\times0.12$	&64 		&0.17			&136--356 
\\ 
WF2c     	&05:35:13.638   &$-05.23.54.94$	&$51\pm2$     	&$236\pm11$  	&$0.50\times0.42$	&128		&0.61			&136--355\\ 
WF3         	&05:35:13.558   &$-05.23.59.08$	&$134\pm3$    	&$188\pm6$ 	&$0.17\times0.13$	&31		&0.49			&136--359 
\\ 
\hline
SF1		&05:35:13.125	 &$-05.24.10.63$	&$5.9\pm0.8$	&$9\pm2$ 	&$-\times-$		&$-$		&0.02			&-\\
SF2a       	&05:35:13.204   &$-05.24.12.57$ 	&$72\pm4$     	&$245\pm16$ 	&$0.47\times0.28$	&13		&0.64			&132--413\\ 
SF3a	&05:35:13.339	 &$-05.24.11.59$	&$15\pm1$     	&$56\pm6$ 	&$0.51\times0.30$	&47		&0.15			&-\\
SF3b       	&05:35:13.409   &$-05.24.11.27$	&$106\pm5$    	&$236\pm17$  	&$0.36\times0.18$	&48		&0.61			&134--411\\ 
SF4a      	&05:35:13.601   &$-05.24.09.13$ 	&$30\pm2$      	&$66\pm5$ 	&$0.42\times0.14$	&92		&0.17			&-\\ 
SF4b       	&05:35:13.631   &$-05.24.09.17$ 	&$56\pm1$    	&$83\pm3$ 	&$0.20\times0.14$	&82		&0.22			&-\\ 
SF4c		&05:35:13.661	 &$-05.24.09.03$	&$7\pm1$    	&$10\pm2$ 	&$0.27\times0.09$	&63		&0.03			&-\\
SF5a      	&05:35:13.735   &$-05.24.07.77$	&$111\pm3$     &$591\pm20$	&$0.58\times0.44$	&74		&1.54			&137--408\\ 
SF5b      	&05:35:13.770   &$-05.24.07.35$	&$23\pm2$     	&$379\pm38$	&$0.98\times0.92$	&143		&0.98			&-\\ 
SF6a       	&05:35:13.903   &$-05.24.08.16$	&$38\pm1$    	&$49\pm2$	&$0.15\times0.12$	&88		&0.13			&-\\ 
SF6b      	&05:35:13.931   &$-05.24.09.45$	&$116\pm4$  	&$280\pm12$	&$0.32\times0.25$	&111		&0.73			&139--409\\ 
\hline
\end{tabular}
\begin{list}{}{}
\item[$^\mathrm{a}$] A lower-case letter in the source labeling indicates different components within a core, as defined in Section 4.1.
\item[$^\mathrm{b}$] Position, peak intensity, flux density, deconvolved size and position angle derived by fitting a 2D Gaussian using the image obtained after combining the extended+compact configurations. Uncertainty in the peak intensity and flux density is given by the fit. Point sources should have deconvolved sizes smaller than $0.10''\times0.05''$.
\item[$^\mathrm{c}$] Masses derived using the flux density given in column (5), and assuming a dust temperature of 25~K, and a dust (+gas) mass opacity coefficient at 1.3~mm of 0.00899~cm$^2$\,g$^{-1}$ (Ossenkopf \& Henning 1994, see main text). The uncertainty in the masses due to the opacity law is estimated to be a factor of 2.
\item[$^\mathrm{d}$]  Other names given in the literature (Smith et al. 2004; Zapata et al. 2005).
\end{list}
}
\end{center}
\label{talma}
\end{table*}

In Table~\ref{talma} we give the  position, peak intensity, flux density, deconvolved size and P.A. of the 31 identified millimeter sources, after performing 2D Gaussian fits (with the task IMFIT in CASA) using the extended+compact configuration image (Fig.\ref{falmaall}). In the table, we also provide an estimate of the mass of gas and dust for each millimeter source, using the flux density given in column (5) of Table~\ref{talma}, and assuming a dust temperature of 25~K, and a dust (+gas) mass opacity coefficient at 1.3~mm of 0.00899~cm$^2$\,g$^{-1}$ (Column 6 of Table 1 of Ossenkopf \& Henning 1994, corresponding to agglomerated dust grains with thin ice mantles at densities $10^6$~\cmt). 
The dust temperature is estimated from observations of the \nh(1,1) and (2,2) transitions by Wiseman \& Ho (1998) of the Orion-KL and OMC-1S regions, and assuming that gas and dust are thermalized in the dense cores. These authors provide an image of the ratio of \nh(2,2) to (1,1) integrated intensities, and for OMC-1S the ratio ranges from 0.5 to 1, corresponding to rotational temperatures from 18 to 33~K (for opacities of the \nh(1,1) transition around 1, Table~2 of Wiseman \& Ho 1998). Thus, we adopt an average temperature of 25 K for all the millimeter sources. Although some of the millimeter sources might be slightly more evolved than others (see Section~4) we adopt the same dust temperature because no big variations are expected given the embedded nature of the OMC-1S region (\eg\ S\'anchez-Monge et al. 2013).
We estimated an uncertainty in the masses due to an uncertainty in the dust opacity of a factor of about two. The derived masses range from 0.02 to 1.5~\mo.

Using the 7~mm data from the VLA in its most extended configuration, we can further study the substructure of the ALMA sources down to $0.09''$ or 40~AU. We identified those 7~mm sources above the 5$\sigma$ threshold.  
The VLA 7~mm 5$\sigma$ detections correspond to a 1.3~mm intensity of $\sim18$~m\jpb, assuming a spectral index of 2. Thus, the ALMA sources fainter than this threshold could not have been detected by the VLA at 7~mm.
The small panels of Figs.~\ref{fvlan}, \ref{fvlaef}, \ref{fvlawf} and \ref{fvlasf} present a zoom of the ALMA sources (above $\sim18$~mJy) in colorscale, with the 7~mm continuum emission overplotted in black contours. 
As can be seen in Fig.~\ref{fvlan}, the ALMA sources N3 and N1b have clear counterparts at 7~mm. In the Eastern Filament, sources EF1, EF2 and EF6 also have 7~mm counterparts (Fig.~\ref{fvlaef}), similar to WF2b and WF3 in the Western Filament (Fig.~\ref{fvlawf}). Finally, in the Southern Filament, SF2, SF3b, SF5 and SF6b have 7~mm counterparts above the 5$\sigma$ threshold (Fig.~\ref{fvlasf}).
After comparison of the 7~mm emission among the different regions/filaments, it is obvious that while the 7~mm counterparts in the northern region, Eastern Filament and Western filament are single sources, an important fraction of the ALMA sources in the Southern Filament split up into several sources at 7~mm.

Table~\ref{tvla} summarizes the properties of the identified 7~mm sources: their peak position, peak intensity, and flux density (integrated inside the 2.5$\sigma$ contour).
In the Table, we also give the peak intensity of the 1.3~cm emission from Zapata et al. (2004). Using these values, and taking into account the positive spectral indices inferred by Zapata et al. (2004) for the 1.3~cm counterparts, a non-negligible part of the 7~mm emission could be contribution from the free-free emission detected at longer wavelengths. Therefore, we cannot disentangle with this information only whether the 7~mm emission comes mainly from free-free or thermal dust emission, and we refrain from estimating masses of gas and dust using the 7~mm emission.

\begin{table}
\caption{Parameters of the sources detected with the VLA at 7~mm in OMC-1S}
\begin{center}
{\small
\begin{tabular}{lccccc}
\noalign{\smallskip}
\hline\noalign{\smallskip}
&\multicolumn{2}{c}{Position$^\mathrm{a}$}
&$I_\mathrm{\nu}^\mathrm{7mm}$~$^\mathrm{a}$
&$S_\mathrm{\nu}^\mathrm{7mm}$~$^\mathrm{a}$
&$I_\mathrm{\nu}^\mathrm{1.3cm}$~$^\mathrm{b}$
\\
\cline{2-3}
Source
&$\alpha (\rm J2000)$
&$\delta (\rm J2000)$
&(mJy)
&(mJy)
&(mJy)
\\
\noalign{\smallskip}
\hline\noalign{\smallskip}
N1b         	&05:35:13.709   &$-05.23.46.88$ 	&$1.91$	&$17.2$				&0.59	\\
N3   		&05:35:14.392   &$-05.23.50.82$	&$1.88$   	&$5.19$	     			&0.78	\\
\hline
EF1         	&05:35:13.876   &$-05.23.57.22$ 	&$1.65$   	&$2.30$	 			&0.55	\\ 
EF2         	&05:35:14.135   &$-05.23.56.68$	&$1.14$   	&$1.49$  				&0.85	\\ 
EF6         	&05:35:14.536   &$-05.23.56.02$	&$0.76$ 	&$2.03$ 				&$<0.21$	\\ 
\hline
WF2b     	&05:35:13.571   &$-05.23.55.80$ 	&$1.32$ 	&$1.75$				&0.57	\\ 
WF3         	&05:35:13.558   &$-05.23.59.07$	&$2.47$  	&$5.37$ 				&1.00	\\ 
\hline
SF2aa      	&05:35:13.203   &$-05.24.12.62$ 	&$0.80$ 	&$4.06$~$^\mathrm{c}$	&0.26	\\ 
SF2ab	&05:35:13.206	 &$-05.24.12.54$	&$0.85$ 	&					&$<0.21$	\\
SF3ba      	&05:35:13.405   &$-05.24.11.33$	&$1.14$  	&$4.13$~$^\mathrm{c}$	&0.78	\\ 
SF3bb      	&05:35:13.411   &$-05.24.11.23$	&$1.27$  	&					&0.78	\\ 
SF5aa      	&05:35:13.737   &$-05.24.07.76$	&$0.93$ 	&$13.1$~$^\mathrm{c}$	&0.15	\\ 
SF5ab      	&05:35:13.742   &$-05.24.07.82$	&$0.96$  	&					&0.15	\\ 
SF5ac      	&05:35:13.742   &$-05.24.07.76$	&$0.84$  	&					&$<0.21$	\\ 
SF5ad      	&05:35:13.746   &$-05.24.07.57$	&$0.69$  	&					&$<0.21$	\\ 
SF6ba      	&05:35:13.928   &$-05.24.09.41$	&$1.09$  	&$2.43$~$^\mathrm{c}$	&0.42	\\ 
SF6bb      	&05:35:13.933   &$-05.24.09.47$	&$0.96$  	&					&0.42	\\ 
\hline
\end{tabular}
\begin{list}{}{}
\item[$^\mathrm{a}$] Position of the peak intensity, peak intensity and flux density integrated within the 2.5$\sigma$ contour. Uncertainty in the peak intensity is the rms noise of the image, 0.13~m\jpb. Units of the peak intensity are m\jpb.
\item[$^\mathrm{b}$] Peak intensity at 1.3~cm from Zapata et al. (2004). For non-detections, an upper limit of 3$\sigma$ is given. For SF3b, SF5 and SF6b the position of the peak of the emission at 1.3~cm lies exactly in between the 7~mm positions of SF3ba and SF3bb, SF5aa and SF5ab, and SF6ba and SF6bb. Thus, we tentatively assigned half of the 1.3~cm peak intensity to each of these 7~mm sources.
\item[$^\mathrm{c}$]  The flux density integrated within the 2.5$\sigma$ contour includes both `a' and `b' sources for SF2, SF3b, and SF6b, and `aa', `ab', `ac' and `ad' sources for SF5. 
\end{list}
}
\end{center}
\label{tvla}
\end{table}

\section{Analysis}

\subsection{Typical separations between the ALMA sources}

\begin{figure}
\begin{center}
\begin{tabular}[b]{c}
    \epsfig{file=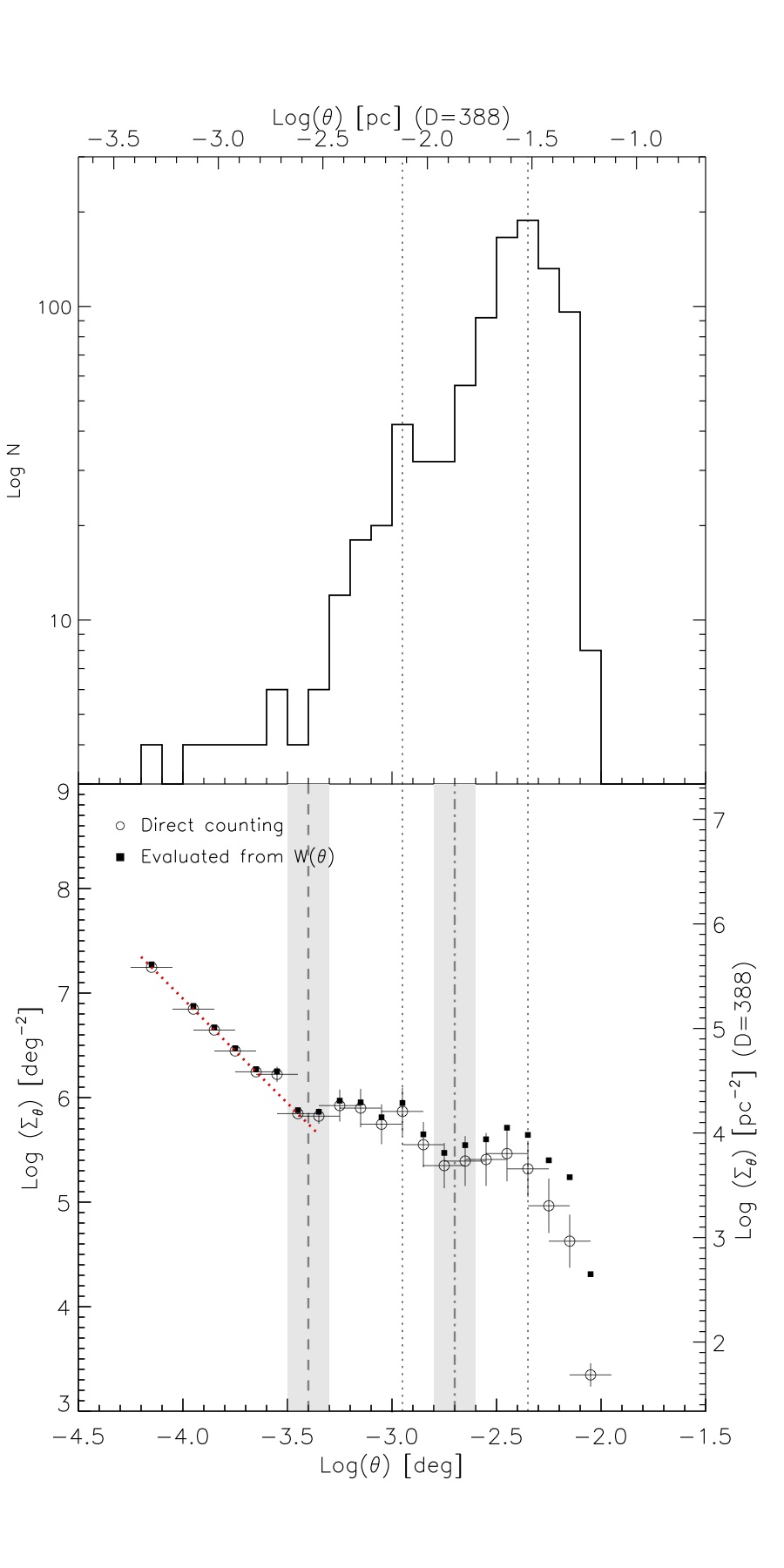, scale=0.27, angle=0}\\
\end{tabular}
\caption{
\emph{Top:} Histogram of separations between the ALMA sources listed in Table~\ref{talma}. The vertical dotted lines correspond to the peaks of the histogram, at $\log{\theta}\sim-2.35$ (1570~AU) and $\log{\theta}\sim-2.95$ (6240~AU).
\emph{Bottom:} Mean surface density of companions vs angular separation. White circles and black squares correspond to the `direct counting' and `$W(\theta)$' (or two-point correlation function) methods, respectively (Section 4.1). The dotted red line corresponds to the power-law fit with index $-1.05$ in the range $-4.5<\log{\theta}<-3.4$. The dot-dashed and dashed vertical lines surrounded by gray bands indicate two observed fragmentation breaks within an estimated confidence interval of 0.1 dex in log$(\theta)$.
}
\label{fseparations}
\end{center}
\end{figure}

In this Section we aim at studying the spatial distribution of the ALMA sources. For this, we first calculated the angular separations, $\theta$, from each to every other source in Table~\ref{talma} using the RA, DEC coordinates. Fig.~\ref{fseparations}-top presents the histogram of separations between the ALMA sources. As can be seen from the figure, the peak of the histogram is at $\log{\theta} \approx-2.35$, corresponding to 6240 AU. This is indicating that the three main filaments in OMC-1S are approximately separated the same projected distance (as can be seen from Fig.~\ref{falmaall}), which would naturally yield that the most common separation between ALMA sources is the typical distance between the filaments. The second peak in the histogram is around $\log{\theta} \approx-2.95$, or 1570 AU, which corresponds to the typical separation between the sources within a filament (this will be further analyzed below).

Second, we calculated the mean surface density of companions (MSDC), $\Sigma _\theta$, for the sources listed in Table~\ref{talma}. 
The MSDC has been widely used to investigate the transition from pairs to groups in distinct scenarios that include young star clusters and dense cores (G\'omez et al. 1993; Bate et al. 1998; Kraus \& Hillenbrand 2008; Rom\'an-Z\'u\~niga et al. 2010, Tafalla \& Hacar 2015). 
In the $\Sigma _\theta$ vs $\theta$ diagram, the binary and clustering regime present a steep negative slope, while a uniform distribution of sources would present a flat slope (\eg\ Bate et al. 1998, Hartmann 2002, Kraus \& Hillenbrand 2008).

Following the prescription by Simon (1997), we took the angular separations $\theta$ and grouped them in 0.1 dex annuli within $-6<\log{\theta}<0$. Then, $\Sigma _\theta$ is computed as the number of elements in each annulus, $N_\mathrm{p}(\theta)$, divided by the area, normalized by the total number of elements. The MSDC is equivalent this way, to the two-point correlation function (TPCF), $W_\theta$, that compares the value of $\Sigma _\theta$ with a uniform, random distribution of points in an area of the same size as our field ($A_\mathrm{m}=10^{-4}\,\mathrm{deg \ }^2$). If $N_\mathrm{r}$ is the number of the randomly distributed points that fall in each of our annuli, then $W_\theta=(N_\mathrm{p}/N_\mathrm{r})-1$. To determine $N_\mathrm{r}$, we averaged the output from $5\times 10^4$ drawings using a simple Monte Carlo routine. Then, $$\Sigma _\theta(\mathrm{TPCF})=( N_\mathrm{p}/A_\mathrm{m})(1+W_\theta),$$ can be compared directly to the value of $\Sigma _\theta$ estimated from direct counting in the anular bins.

The results of this calculation are shown in Fig.~\ref{fseparations}-bottom. The figure shows that $\Sigma _\theta$ is anticorrelated to $\theta$  from $-4.5<\log{\theta}<-3.4$ (about 44 to 556 AU), then the distribution appears to flatten out to $\log{\theta} \approx-3$ (about 1397 AU), followed by another short fall and a flattening or small bump from $-3$ to $-2.7$ and from $-2.7$ to $-2.5$, respectively. The values of the MSDC and the TPCF are very consistent down to this last value, but above it, the function shows an abrupt decrease, indicating where the angular separation reaches the field size. 
Two aspects are remarkable: one is that the linear fit in the $-4.5<\log{\theta}<-3.4$ range, or 44--556~AU, has a well defined slope of $-1.05$, consistent with ordered fragmentation and similar to the power-law index expected for the regime dominated by the distribution of binaries (\eg\ Bate et al. 1998).
The linear fit and the slope near $-1$ are also similar to values obtained by Simon et al. (1997) in Orion Trapezium and by Kraus \& Hillenbrand (2008) in Taurus and Upper Sco over a similar range of scales.
The second aspect is the two observed breaks, with respect to the slope $-1$, at $\log{\theta}=-3.4$ and $-2.7$ (556 and 2787 AU respectively), followed by a flattening. Both breaks are marked with vertical dashed lines in Fig.~\ref{fseparations}-bottom. While the second break is more doubtful because there are only very few points in the negative slope side, the break at $\log{\theta}=-3.4$ or 556~AU presents a clear flattening with respect to the power law fit. 
Since a flattening is expected for a uniform distribution of sources, the breaks indicate the size scales of the binary regime and source surface density enhancements.
Therefore, the spatial scale at 556~AU could be associated with the size of cores which fragment and yield to the formation of binary/multiple systems.


\subsection{Cores defined as groups of ALMA sources}

In the previous section a particular spatial scale associated with the size of binary and multiple systems was found at $\sim556$~AU. We use here such a spatial scale to define a `core', as the grouping of those millimeter sources for which the distance between each member of the group and at least another member is $<556$~AU. To illustrate this more clearly, in the Appendix we present the ALMA images with circles of 556~AU of radius centered on each ALMA source listed in Table~\ref{talma}. From those figures, it is straightforward to see which sources are separated $<556$~AU, and thus which sources would belong to the same `core' (marked in the figure with the same color code). Because our data present indeed an excess of source surface density at such scale, the cores are well defined and do not extend indefinitely through a filament. According to our definition, cores are structures of $\sim1100$~AU of diameter which might harbor one single millimeter source or several millimeter sources. By doing this, we identified 19 cores, listed in Table~\ref{tNmm}.

We calculated the geometric center of each core by performing the average of the positions of the ALMA sources belonging to the same core (Table~\ref{tNmm}), and estimated the average (standard deviation) separation between cores within each filament: 1310(460)~AU for the Eastern Filament,  1280(300)~AU for the Western Filament, and 1140(350)~AU for the Southern Filament. These average separations are consistent with the secondary peak at 1570 AU of the histogram of separations shown in Fig.~\ref{fseparations}-top.


\subsection{Fragmentation level vs density}

The ALMA and VLA observations have revealed the structure of the millimeter emission in OMC-1S, at spatial scales spanning almost two orders of magnitude, from 2200~AU (corresponding to the Largest Angular Scale of the ALMA data) down to $\sim40$~AU (corresponding to the angular resolution of the VLA data). This allowed us to identify three extended filamentary structures, with 19 cores embedded in them, as well as further sources within these cores. 

In the following, we define the `fragmentation level' of each core as the number of closed contours above the detection threshold at each wavelength (12$\sigma$ at 1.3~mm, and 5$\sigma$ at 7~mm), within a region of 1100~AU of diameter, as this is approximately the core size. This is similar to the definition used in Palau et al. (2013, 2014, 2015), but in these works the fragmentation was assessed within a region of 0.1~pc, while in the present work we assess the fragmentation in a region about one order of magnitude smaller. In the case of no closed contours but the single Gaussian fit clearly requiring an additional Gaussian (see Section~3), we counted each Gaussian as a separate source (this is the case of N1 and SF5 in the ALMA data). In the cases where the ALMA source has a clear 7~mm counterpart, we have counted them as one single source.  
Our VLA+ALMA dataset allows us to study the fragmentation properties in OMC-1S down to spatial resolutions of $\sim40$~AU and for mass sensitivities of $\sim0.05$~\mo\ (estimated using a flux density of 18~mJy at 1.3~mm and the same assumptions of Table~\ref{talma}; the flux density threshold corresponds to the extrapolation of the 7~mm sensitivity to 1.3~mm, see Section~3). For this reason, the fragmentation in those cores whose peak intensity is below 18 mJy has not been studied down to 40~AU but to 74~AU, and thus its fragmentation level is considered as a lower limit. This is the case of 6 out of the 19 cores.

In Table~\ref{tNmm}, we list the 19 ALMA cores and their fragmentation level within 1100~AU. From that table, it is obvious that all of the cores in the Southern Filament present a fragmentation level higher than all the cores in the Eastern Filament (we are not taking into account here the cores for which we only have lower limits in the fragmentation level). 
To quantify whether the differences in fragmentation between the Southern and Eastern Filaments are significant, we calculated the probability of drawing three `ones' (fragmentation level of the three cores in the Eastern Filament) from a Gaussian distribution with the mean and standard deviation equal to that of the Southern Filament, which are 3.200 and 1.095, respectively. This yields a probability of drawing one `one' of 0.041. Then, the probability of drawing three ones as in the Eastern Filament is $\sim0.0071$\%, which is really small. This indicates that the fragmentation level in the Southern Filament is significantly higher than in the Eastern Filament.

\begin{figure}
\begin{center}
\begin{tabular}[b]{c}
    \epsfig{file=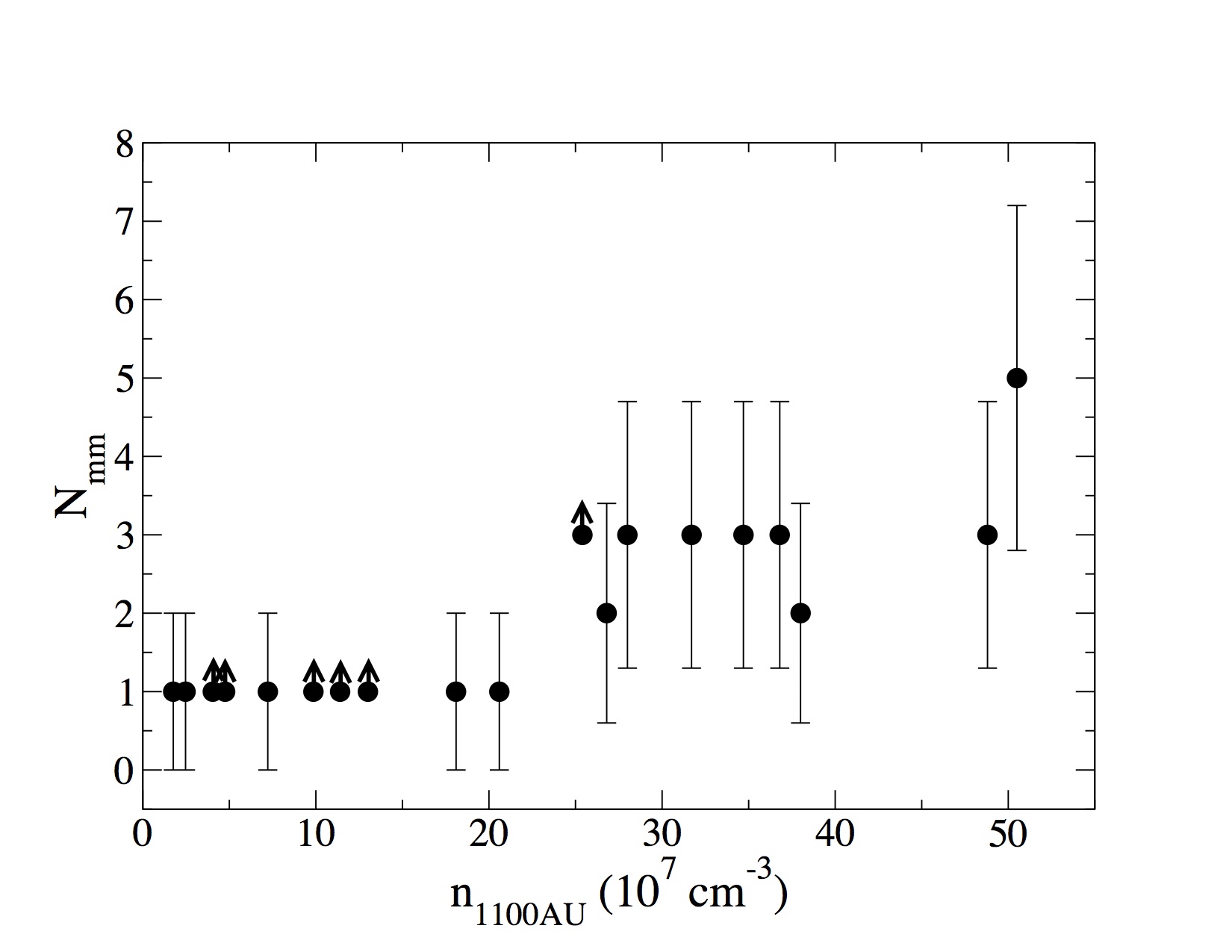, scale=0.15, angle=0}\\
    \epsfig{file=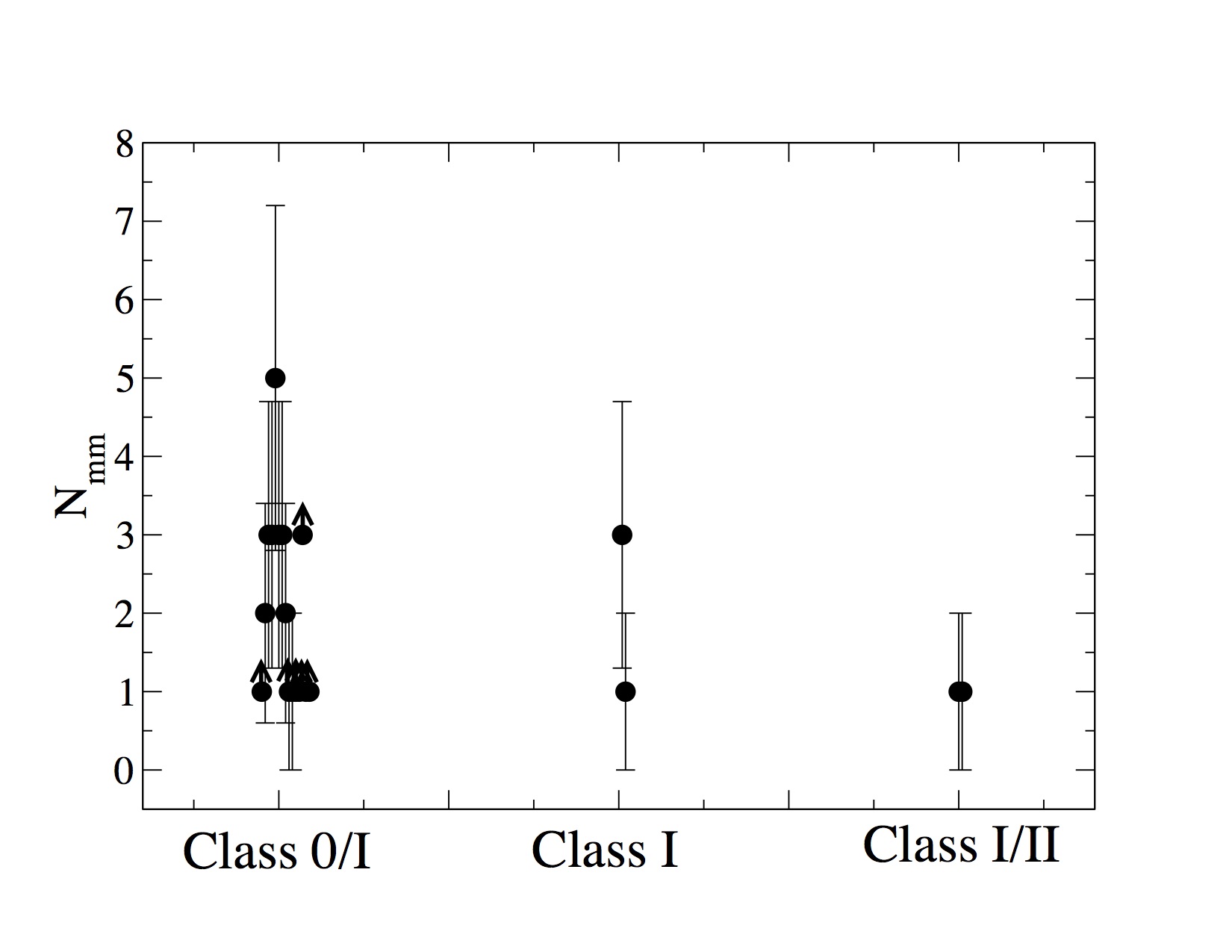, scale=0.15, angle=0}\\
\end{tabular}
\caption{
\emph{Top:} Fragmentation level vs density (of H$_2$ molecules) enclosed within 1100~AU. 
\emph{Bottom:} Fragmentation level vs tentative evolutionary class.
}
\label{fNmm}
\end{center}
\end{figure} 

An ingredient which could be playing a role in the early fragmentation of the cores is their density structure. This was studied by Palau et al. (2014), who find that the fragmentation level in a sample of massive dense cores of $\gtrsim0.1$~pc seems to be tightly related to the density of the core within the diameter of 0.1~pc, which in turn depends on the density profile and the central density of the core. With this in mind, we used an approximate method to test in our data the effect of the density structure of the core, assuming that the temperature structure does not change significantly from one core to the other and thus that the differences in the intensity profiles mainly reflect the differences in density profiles. 
For each core, we estimated the mass (in the ALMA image including the extended and compact configurations, and following the same assumptions of Section~3) enclosed within a given diameter, which we take to be 1100~AU, and then converted this into density. By doing this, if the density profile is rather flat we will obtain a high mass within the given diameter, while if the density profile is highly concentrated, the mass within the same diameter will be smaller (assuming similar central densities). 
In Fig.~\ref{fNmm}-top we plot the fragmentation level of each core against the density within 1100 AU. 
The figure shows that the higher the density, the higher the fragmentation level\footnote{We would like to emphasize that the density averaged over all the core is not correlated with the fragmentation level because this average density is not taking into account the density distribution. What is well correlated with the fragmentation level is the density averaged within a given radius, which takes into account the central density and the density profile of the core.}. Actually, the average density within 1100~AU for the fragmenting cores is $36\times10^{7}$~\cmt, while the average density within 1100~AU for the non-fragmenting cores is about a factor of 4 smaller, $9.3\times10^{7}$~\cmt.
%


\subsection{Fragmentation level vs evolutionary stage}

\begin{figure*}
\begin{center}
\begin{tabular}[b]{c}
    \epsfig{file=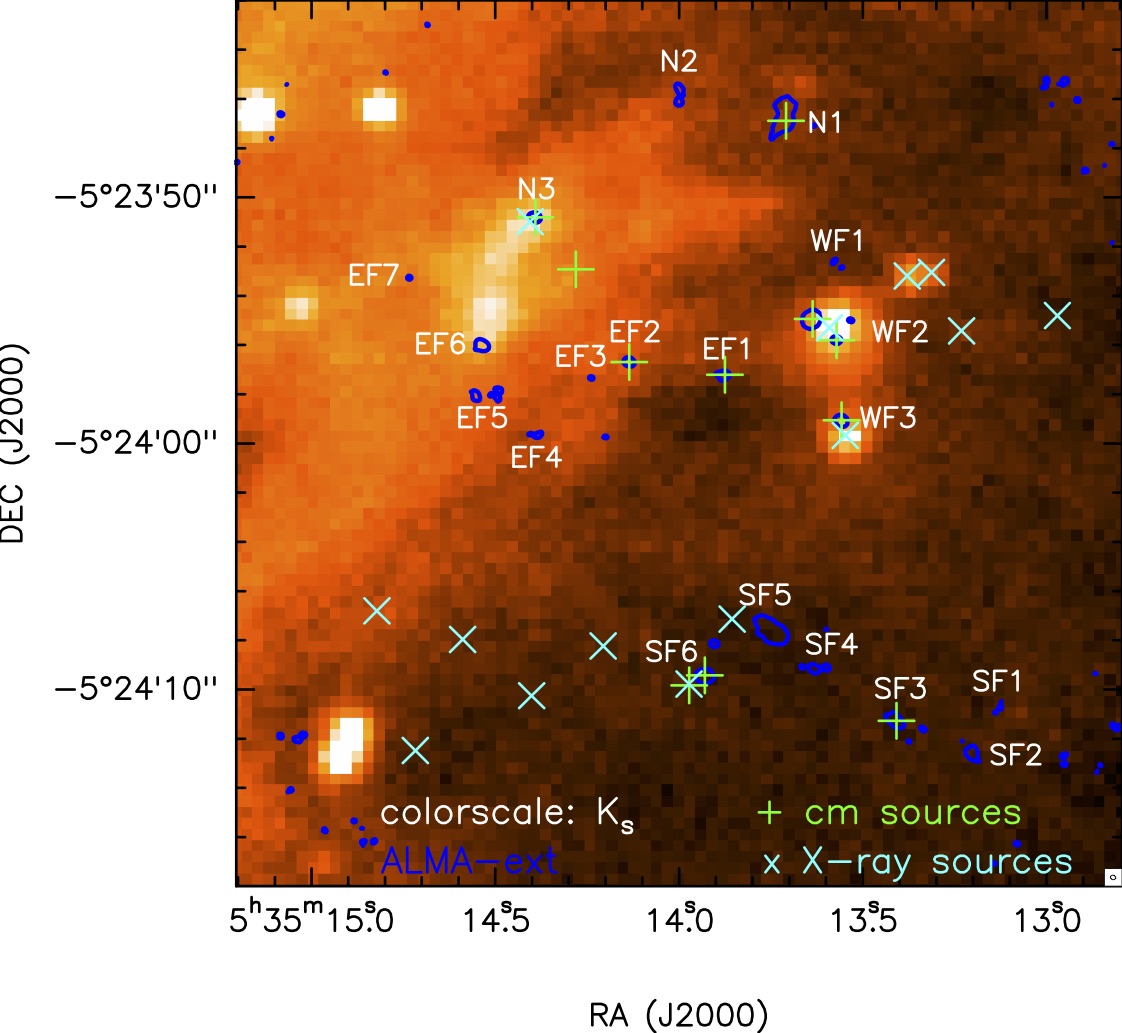, scale=0.4,angle=0}\\
\end{tabular}
\caption{Colorscale: CAHA $K_\mathrm{s}$ band. Blue contours: ALMA 1.3~mm continuum emission using only the extended configuration. Contours correspond to the 12 times the rms noise of the map, 0.45~m\jpb. The 19 identified ALMA cores, where we have studied fragmentation, are labeled in white, and the beam of the ALMA image is shown in the bottom right corner. Green plus signs: 1.3~cm sources from Zapata et al. (2004); Turquoise crosses: X-ray sources from Rivilla et al. (2013).}
\label{fcaha}
\end{center}
\end{figure*}

We have explored the possibility that the high fragmentation in the Southern Filament is related to the density of each core, and find indications of a trend between fragmentation and density. Another factor which could also determine the fragmentation level is the evolutionary stage. 
Recent studies in the literature are suggestive of a trend of multiplicity fraction\footnote{At the small spatial scales where we have studied fragmentation ($\sim500$~AU of radius), one could refer to the sources belonging to the same core as forming a `multiple system'. However, from our data alone it is not possible to assure the protostellar nature of all the detected millimeter sources, and it is neither possible to study whether the sources belonging to each core are gravitationally bound or not. For these reasons, we refrain from calling our study a `multiplicity study' and prefer to call it a `fragmentation study'.} decreasing with evolutionary stage, from Class 0 to Class I (\eg\ Connelley et al. 2008, Chen et al. 2013,  Duchene \& Kraus 2013, Tobin et al. 2016). 
%
Such a trend is expected if fragmentation takes place bound to the gas potential and the potential is dominated by one central object, naturally leading to the decrease of distances between the fragments (\eg\ Sadavoy \& Stahler 2017).
It is also expected from dynamical evolution, since three-body interactions tend to destroy wide binary systems (\eg\ Kroupa et al. 1999; Kaczmarek, Olczak, \& Pfalzner 2011; Duchene \& Kraus 2013; Li et al. 2016).

To explore this possibility, the most appropriate method would be to build the Spectral Energy Distribution (SED) for each core. However, to do this we would need mid-infrared and far-infrared information for each core at an angular resolution better than $\sim2''$, and this is not available for the ALMA cores in OMC-1S. We crossed our list of cores with the Orion \emph{Spitzer} and \emph{Herschel} catalogs of Megeath et al. (2012) and Furlan et al. (2016), and only found identified \emph{Spitzer}/IRAC point sources corresponding to N3 and WF2b, and no sources in \emph{Spitzer}/MIPS or \emph{Herschel}/PACS catalogs. The sources N3 and WF2b are also detected in the mid-infrared observations of Smith et al. (2004), together with EF1, EF6 and WF3. With a total of 4 cores with mid-infrared detections, it is definitely not possible to study any trend of fragmentation with SED properties. Instead, we have collected data for all the cores at different wavelengths, so that we may have a rough and very approximate idea of the possible evolutionary stage of each core. 

On one hand, in Fig.~\ref{fcaha} we present a CAHA $K_\mathrm{s}$ image of the OMC-1S region. To obtain this near-infrared image, we used the Calar Alto Public Archive\footnote{http://www.caha.es/caha-public-archives.html} and download a set of near-infrared images taken between 2011 December 13 and 15, with the OMEGA-2000 Camera on the 3.5m telescope at Calar Alto Observatory. We processed the raw images along with a set of proper calibration files using our reduction pipelines\footnote{In a nutshell, our pipelines consist of IRAF and Fortran routines that provided a  two-pass sky subtraction reduction, centroid corrected dither registering, a $<0.2\arcsec$ rms astrometric solution, as well as aperture and PSF photometry catalogs for all detectable sources.}, in a sequence similar to that described by Rom\'an-Z\'u\~niga et al. (2015).  
As can be seen in the figure, some of the ALMA cores (overplotted in blue contours) are detected in the infrared, and these are mainly located in the Western and Eastern Filaments. In addition, we searched for \emph{Chandra} X-ray counterparts to the ALMA cores (Rivilla et al. 2013). Finally, we also studied the 1.3~cm emission associated with the cores, following Zapata et al. (2004), together with their respective spectral indices. 
We summarize all these properties in Table~\ref{tNmm} and Fig.~\ref{fcaha}.

After compiling all these data for the ALMA cores, we tentatively assigned an evolutionary stage for each core. We assigned the stage of Class 0/I for the most embedded sources, and Class I/II for the more evolved objects probably having only a faint envelope. Intermediate cases would be assigned the Class I stage. 
We based our classification on the following criteria: 
i) near-infrared emission and X-ray emission are usually associated with young stellar objects dominated by a disk (no envelope, \eg\ Rivilla et al. 2013; Povich et al. 2016), and detection in at least one of these signposts, and association with HH objects (no molecular outflows) would suggest a Class I/II nature; 
ii) if a source is detected in all the signposts of Table~\ref{tNmm}, with the outflow being molecular, we consider it to be in an intermediate stage (\ie\ Class I);
iii) sources driving molecular outflows (traced by SiO, CO, or water masers) and being associated with thermal centimeter emission (positive spectral index between 3.6 and 1.3~cm, Zapata et al. 2004), with no detection in the infrared or X-rays are typically deeply embedded sources (\eg\ Liu et al. 2014, Pech et al. 2016), suggesting a Class 0/I nature; 
iv) if a source has not been detected in any of the signposts listed in Table~\ref{tNmm}, we consider it as deeply embedded (\ie\ Class 0/I), since the extreme compact nature of the dust emission suggests that at least an hydrostatic core should have already been formed.

In Fig.~\ref{fNmm}-bottom we plot the fragmentation level against this tentative evolutionary stage, which could be suggestive of a decreasing trend.
However, it should be kept in mind that the statistics used to build Fig.~\ref{fNmm}-bottom are extremely poor (essentially because most of the sources seem to be in the Class 0/I stage), and that in the earliest stage there is a wide spread of fragmentation levels: from the lowest possible (around 1 source) up to the highest (5 sources). 
Therefore, to draw firmer conclusions regarding a possible trend between the fragmentation level and evolutionary stage, better statistics especially in the later stages should be compiled.


\begin{table*}
\caption{Summary of properties of the cores embedded  in OMC-1S}
\begin{center}
{\small
\begin{tabular}{lcccccccccc}
\noalign{\smallskip}
\hline\noalign{\smallskip}
&\multicolumn{2}{c}{Position$^\mathrm{a}$}
&$n_\mathrm{1100AU}$$^\mathrm{b}$
&&&&&&&Tentative
\\
\cline{2-3}
Core
&$\alpha (\rm J2000)$
&$\delta (\rm J2000)$
&($\times10^7$~\cmt)
&$N_\mathrm{mm}$$^\mathrm{b}$
&IR?~$^\mathrm{c}$
&X-rays?~$^\mathrm{c}$
&Outf?~$^\mathrm{c}$
&HMC?~$^\mathrm{c}$
&cm?~$^\mathrm{c}$
&Class~$^\mathrm{c}$
\\
\noalign{\smallskip}
\hline\noalign{\smallskip}
N1         	&05:35:13.713   &$-05.23.46.77$	&49	&3				&N	&N	&SiO		&Y	&Y	&0/I	\\
N2       	&05:35:13.998   &$-05.23.45.86$	&27	&2				&N	&N	&N		&N	&N	&0/I	\\
N3   		&05:35:14.392   &$-05.23.50.82$	&1.8	&1 				&Y	&Y	&HH		&N	&Y	&I/II	\\
\hline
EF1         	&05:35:13.878   &$-05.23.57.21$ 	&2.5	&1 				&N	&N	&SiO?	&Y	&Y	&0/I\\ 
EF2         	&05:35:14.135   &$-05.23.56.69$	&7.2	&1 				&N	&N	&SiO?	&Y	&Y	&0/I\\ 
EF3		&05:35:14.238   &$-05.23.57.38$	&4.8	&$\geqslant1$		&N	&N	&N		&N	&N	&0/I\\
EF4		&05:35:14.386   &$-05.23.59.64$	&13	&$\geqslant1$		&N	&N	&N		&N	&N	&0/I\\
EF5		&05:35:14.520   &$-05.23.57.97$	&25	&$\geqslant3$		&N	&N	&N		&N	&N	&0/I\\
EF6         	&05:35:14.535   &$-05.23.56.02$	&21	&1  				&Y	&N	&OOS	&N	&N	&I/II\\ 
EF7		&05:35:14.734   &$-05.23.53.28$	&4.0	&$\geqslant1$		&N	&N	&N		&N	&N	&0/I\\
\hline
WF1		&05:35:13.578   &$-05.23.52.50$	&9.9	&$\geqslant1$		&N	&N	&N		&N	&N	&0/I\\
WF2$^\mathrm{d}$&05:35:13.580   &$-05.23.55.25$ &32	&3   			&Y	&Y	&SiO		&Y	&Y	&I\\ 
WF3         	&05:35:13.558   &$-05.23.59.08$	&18	&1    				&Y	&Y	&CO		&Y	&Y	&I\\ 
\hline
SF1		&05:35:13.125   &$-05.24.10.63$	&11	&$\geqslant1$		&N	&N	&N		&N	&N	&0/I\\
SF2		&05:35:13.204   &$-05.24.12.57$	&38	&2   				&N	&N	&N		&N	&N	&0/I	\\ 
SF3		&05:35:13.374   &$-05.24.11.43$	&37	&3				&N	&N	&N		&Y	&Y	&0/I\\
SF4		&05:35:13.631   &$-05.24.09.11$     	&35	&3   				&N	&N	&mas	&Y	&N	&0/I	\\ 
SF5 		&05:35:13.752   &$-05.24.07.56$    	&50	&5    				&N	&N	&SiO		&N	&N	&0/I	\\ 
SF6         	&05:35:13.917   &$-05.24.08.80$	&28	&3    				&N	&Y	&mas	&Y	&Y	&0/I	\\ 
\hline
\end{tabular}
\begin{list}{}{}
\item[$^\mathrm{a}$] Core coordinates adopted as the geometric center of the sources belonging to the core.
\item[$^\mathrm{b}$] $n_\mathrm{1100AU}$ is the density of H$_2$ molecules within 1100~AU (of diameter), obtained after measuring the flux density in a region of 1100~AU of diameter ($\sim3''$), and converting this into a mass, adopting the same assumptions as in Table~\ref{talma}. $N_\mathrm{mm}$ is the number of millimeter sources within 1100~AU including both ALMA and VLA data. Lower limits correspond to the ALMA sources which are too faint to be detected with the 7~mm VLA observations.
\item[$^\mathrm{c}$] References for compact infrared emission (IR): Smith et al. (2004) and this work; X-ray emission: Rivilla et al. (2013); Outflows (Outf): in case of outflow detection, we give the outflow tracer: CO (Zapata et al. 2005), SiO (Zapata et al. 2006), HH (Herbig-Haro object, Smith et al. 2004), OOS: region from which at least six HH objects originate, with coordinates (J2000): R.A. 05h35m14.56s, Dec. $-05:23:54$, with a position error of  $\pm1.5''$, O'Dell \& Doi 2003), mas: water masers from Zapata et al. (2007); Hot Molecular Core (HMC): Zapata et al. (2007) and Palau et al., in prep.; 1.3~cm centimeter emission (cm): Zapata et al. (2004). 
The Tentative Class assigned to each core has been estimated from a global assessment of the previous properties (see Section~4). 
\item[$^\mathrm{d}$] For the case of core WF2, we give the properties associated with the strongest millimeter source, WF2b, but note that WF2c is also associated with centimeter emission.
Regarding the hot molecular core, this is associated with WF2c.
\end{list}
}
\end{center}
\label{tNmm}
\end{table*}

\section{Discussion}

In the previous section we studied the fragmentation level of the OMC-1S cores, and showed that most of the cores of the Southern Filament present a high fragmentation level compared to the cores of the Eastern filament. We studied whether this could be related to the evolutionary stage of the cores but cannot infer any conclusion because most of the cores seem to be in a similar, deeply embedded, evolutionary stage, with only 4 cores out of 19 having hints of being in later evolutionary stages. 
Thus, it remains to be understood which process sets the pristine fragmentation level of a core. OMC-1S is an excellent region to study this because of the high number of detected cores with ALMA, and because most of these cores seem to harbor young stellar objects at very early evolutionary stages (Table~\ref{tNmm}). 

There are several processes that have been proposed in the literature to play a role in setting the pristine fragmentation of a core, such as the magnetic field (\eg\ Boss et al. 2004; V\'azquez-Semadeni et al. 2005; Hennebelle et al. 2011, Commer\c con et al. 2011; Fontani et al. 2016), turbulent support (\eg\ Zhang et al. 2009, 2015; Wang et al. 2011, 2014; Pillai et al. 2011, Lu et al. 2015), or the rotational-to-gravitational energy (\eg\  Arreaga-Garcia 2017, Lim et al. 2016).
Recently, Palau et al. (2014, 2015) have studied the aforementioned processes in a sample of 19 massive dense cores and found that the fragmentation level within cores of 0.1~pc is best correlated with the density structure of the cores, rather than with the non-thermal velocity dispersion (Palau et al. 2015) or even the rotational-to-gravitational energy (Palau et al. 2014). Interestingly, Lee et al. (2015) also find no clear correlation of the fragmentation level with non-thermal velocity dispersion or rotational-to-gravitational energy in L1448N, but a good correlation of higher fragmentation with density. Other recent observational works supporting the crucial role of the density on the star formation history of the cores are those of Samal et al. (2015), Sharma et al. (2016), and Figueira et al. (2017). 
All these works favor pure thermal Jeans fragmentation, as in this case, the higher the density of the core, the smaller the Jeans mass and thus the higher number of fragments is expected. In the previous section we have studied this in OMC-1S, and find again an increasing trend of higher fragmentation with density within 1100~AU, very consistent with the previous works but now at scales about one order of magnitude smaller.

If Jeans fragmentation is at work, we should find masses of the sources comparable to the Jeans mass, and separations comparable to the Jeans length. If we take an average density within 1100~AU for the fragmenting cores of about $\sim36\times10^7$~\cmt\ (Section~4.3), and a temperature of 25~K, the Jeans mass is $\sim0.04$~\mo\ (following Palau et al. 2015), while the Jeans length is 360~AU (following Bontemps et al. 2010). For our ALMA data, the mass sensitivity is 0.018~\mo\ (for the 12$\sigma$ detection threshold), fully sensitive to the Jeans mass, and for the VLA data the mass sensitivity is $\sim0.2$~\mo\ (for the 5$\sigma$ detection threshold and in the case of thermal dust emission). Similarly, the Jeans length is fully resolved by our ALMA (and VLA) data ($\lesssim100$~AU). Among the 31 sources detected with ALMA, almost half of them have masses below $\sim0.1$~\mo\ (Table~\ref{talma}), only a factor of 2.5 larger than the Jeans mass. And we should still keep in mind that the masses reported in Table~\ref{talma} should be an upper limit to the true mass of the source because we have calculated the masses using the flux density, and thus including extended emission around the sources, which implies in most cases a factor of 3--10 larger flux density. 

Regarding the Jeans length, this is consistent with the spatial scale of the binary regime found in our ALMA data from the MSDCs analysis presented in Section 4.1, of $\sim 550$~AU.
We can also compare the typical separations of cores along the filaments in OMC-1S to the Jeans length expected for an isothermal, infinite, self-gravitating filament, $3.94\frac{c_\mathrm{s}^2}{G\Sigma_0}$ (Larson 1985, Hartmann 2002), where $c_\mathrm{s}$ is the sound speed, $G$ the gravitational constant, and $\Sigma_0$ the surface density of the filament. Assuming a temperature of 25~K, and a surface density around 2~g\,cm$^{-2}$ (corresponding to the cores in Table~3 with smaller densities), the Jeans length is 1780~AU. This is in full agreement with the measured separations of cores along the filaments, ranging from 1100 to 1300~AU, the secondary peak of the histogram of separations, 1570~AU, and it is only slightly smaller than the second break on the MSDC curve, 2900~AU. It should be kept in mind that the observed separations have not been corrected for projection effects and that these are lower limits. Therefore, both the Jeans mass and the Jeans length seem to be in reasonable agreement with thermal Jeans fragmentation.


This suggests two level Jeans-like fragmentation, as found by Teixeira et al. (2016) in the Orion Molecular Cloud 1 northern filament. These authors study the role of turbulence and magnetic field and find that they cannot explain the observed Jeans lengths, and suggest that thermal Jeans fragmentation is taking place simultaneously with the local collapse of the clumps.
Although the spatial scales where Teixeira et al. (2016) find fragmentation, 2500 and 12400~AU, are larger than the scales studied in this work, the fact that two-level fragmentation is found in both nearby regions suggests that hierarchical fragmentation is a usual process at least in the Orion molecular clouds. This is also supported by a recent work studying fragmentation in the Integral Shaped Filament down to $\sim1000$~AU, where a hierarchical, two-mode fragmentation is also found (Kainulainen et al. 2017).

We would like to remark however that, although it seems that the fragmentation properties in OMC-1S can be explained through pure thermal Jeans fragmentation, a recent work studying multiplicity in the Perseus cloud finds that systems with separations of around $\lesssim200$~AU should have formed through disk fragmentation, while systems with separations $\gtrsim1000$~AU should have formed through thermal fragmentation (\eg\ Tobin et al. 2016). However, this behavior seems to be different from the Taurus cloud (Kraus et al. 2011), and should not necessary apply to Orion. For the case of Orion, fragmentation has been studied on spatial scales larger than 1000~AU and does not cover the `disk fragmentation' domain (\eg\ Kainulainen et al. 2017), preventing from drawing conclusions regarding this model.
Thus the mechanism of formation of the closest systems in OMC-1S (such as SF2, SF5, SF6) should be studied in more detail in order to test the disk fragmentation scenario for these cases.

Finally, a note regarding nomenclature. There is currently some confusion in the literature with the nomenclature used to describe the fragmentation processes, especially those including the role of turbulence. For example, some authors call `turbulent fragmentation' to the process yielding fragments with masses larger than the thermal Jeans mass, \ie\ the Jeans masses are estimated using the Jeans criterion but replacing the factor related to temperature by a factor related to the non-thermal line width (equations 2, 4 and 5 of Palau et al. 2015). This approach presents some limitations, in particular the non-thermal line width could be due to processes different from internal turbulence (\eg\ infall). The data presented in this work can be reproduced without this turbulent support and thus there is no need to invoke turbulent support at these small ($\sim1000$~AU) scales. There are also a number of authors who refer to `gravoturbulent fragmentation' to the fragmentation taking place in a self-gravitating turbulent medium, where the density is determined by the enhancements created by turbulence (\eg\ Padoan \& Nordlund 2002; Schmeja et al. 2004; Fisher 2004; Goodwin et al. 2004; Hennebelle \& Chabrier 2008, Offner et al. 2010; Kirk et al. 2017). In this scenario, the Jeans mass estimation is reduced to a pure thermal Jeans mass at the scales we consider in this study. Our data are thus fully consistent with this gravoturbulent fragmentation, what we call here `pure thermal Jeans fragmentation'.

\section{Conclusions}

With the aim of studying the fragmentation properties down to spatial scales of multiple systems, we conducted 1.3~mm ALMA observations towards OMC-1S, with an angular resolution of $\sim0.2''$, corresponding to spatial scales of $\sim74$~AU, and a sensitivity 2 orders of magnitude better than previous works. In addition, these observations have been complemented with 7~mm VLA observations at $\sim0.1''$ of angular resolution. Our main conclusions can be summarized as follows:

\begin{itemize}

\item[-] The deep 1.3~mm ALMA continuum image reveals 3 filamentary structures in OMC-1S (the Eastern, Western and Southern Filaments), and 31 millimeter continuum sources, most of them embedded within the filaments. The complementary but less sensitive 7~mm observations reveal further substructure within the strongest ALMA sources. This is particularly important in the Southern Filament, where almost all the ALMA cores split up into several sources.

\item[-] A Mean Surface Density of Companions analysis suggests that there are two characteristic spatial scales in OMC-1S: one at 556~AU, and the other at 2900~AU, indicative of a two-level fragmentation process. We used the first spatial scale to define the boundary of `cores' within filaments, and find 19 cores, which consist of groups of several millimeter sources, or single sources.

\item[-] Taking into account both the ALMA and VLA data, a fragmentation level has been assigned to each `core'. There is a correlation of fragmentation level with the density of the core within 1100 AU, as found in previous works studying fragmentation at one order of magnitude larger spatial scales, suggesting that fragmentation at the earliest stages and within spatial scales of $\sim1100$~AU is following a thermal Jeans process.

\item[-] Finally, by compiling near-infrared, X-ray, centimeter, and outflow information for all the ALMA cores, a tentative evolutionary stage is assigned to each core. However, because most of the sources are found to be deeply embedded, no clear conclusions can be drawn regarding a possible trend of fragmentation level with evolutionary stage.

\end{itemize}

In summary, considering that the fragmentation level within 1100 AU can be regarded as a first approach to the multiplicity of the core, the deep ALMA data presented here, together with the VLA data, seem to suggest that multiple systems with separations $>40$~AU are formed through the same mechanisms that govern the fragmentation process at larger scales. This result deserves further investigation in order to disentangle the role of disk fragmentation in the formation of multiple systems in OMC-1S, something which definitely requires kinematic information from molecular gas tracers.

\acknowledgments
\begin{small}
The authors are grateful to the anonymous referee for insightful suggestions, which have improved the quality of the paper.
A.P. is grateful to Eric Keto and Paula Teixeira for thoughtful discussions, and to V\'ictor Rivilla for sharing the X-ray source catalog.
A.P. and C.R.Z. acknowledge financial support from UNAM-DGAPA-PAPIIT IA102815 and UNAM-DGAPA-PAPIIT IN 108117 grants, M\'exico.
L.Z. acknowledges financial support from DGAPA, UNAM, and CONACyT, México.
R.E. acknowledges support by the Spanish MINECO grant AYA2014-57369-C3 (cofunded with FEDER funds) and MDM-2014-0369 of ICCUB (Unidad de Excelencia `Mar\'ia de Maeztu'). 
G.B. acknowledges the support of the Spanish Ministerio de Econom{\'i}a y Competitividad (MINECO) under the grant FPDI-2013-18204. 
R.E., G.B., and J.M.G. are supported by the Spanish MICINN grant AYA2014-57369-C3. 
%
A.F. thanks the Spanish MINECO for funding support from grant AYA2016-75066-C2-2-P and ERC under ERC-2013-SyG, G. A. 610256 NANOCOSMOS.
This paper makes use of the following ALMA data: ADS/JAO.ALMA\#2015.1.00865.S. ALMA is a partnership of ESO (representing its member states), NSF (USA) and NINS (Japan), together with NRC (Canada), NSC and ASIAA (Taiwan), and KASI (Republic of Korea), in cooperation with the Republic of Chile. The Joint ALMA Observatory is operated by ESO, AUI/NRAO and NAOJ.
This research has made use of the VizieR catalogue access tool, CDS, Strasbourg, France. The original description of the VizieR service was published in A\&AS 143, 23.
The authors made use of the GILDAS software (\url{http://www.iram.fr/IRAMFR/GILDAS}) and acknowledge the efforts of the GILDAS team to keep it up to date.
\end{small}



\appendix

In this Appendix we present the identification of each core from the ALMA data, as a result from the Mean Surface Density of Companions analysis presented in Section 4.1. Each core corresponds to a group of ALMA sources or to a single ALMA source. ALMA sources belonging to the same core are indicated with plus signs and circles of the same color.

\begin{figure}
\begin{center}
\begin{tabular}[b]{c}
    \epsfig{file=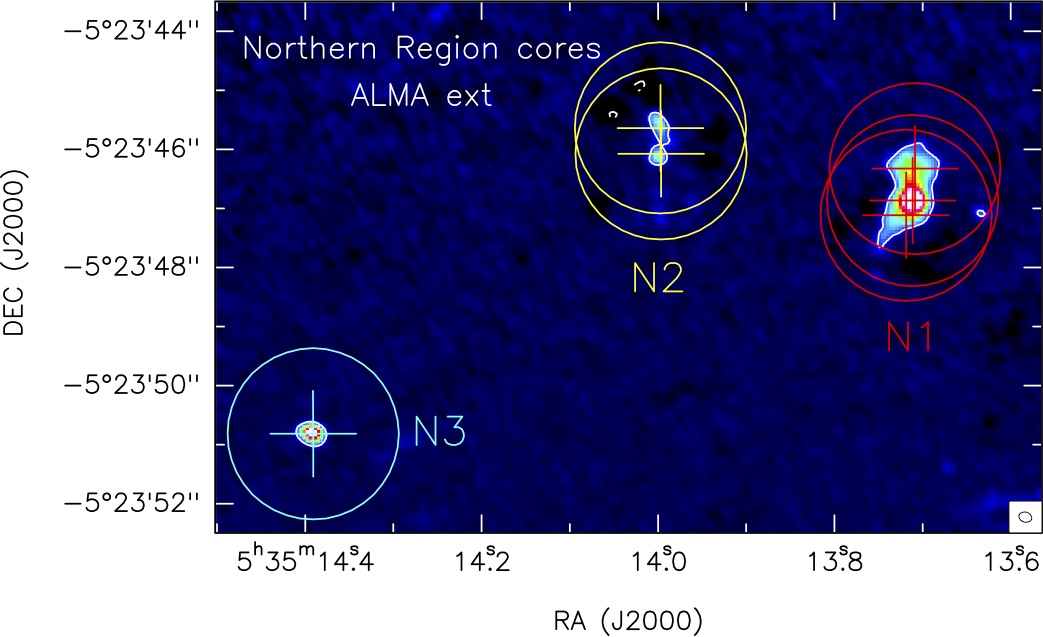, scale=0.4, angle=0}\\
\end{tabular}
\caption{Cores N1, N2 and N3 identified in the Northern region. Plus signs correspond to the positions of the ALMA sources (listed in Table~\ref{talma}), and the circles mark the distance of 556~AU with respect to each source.}
\label{fseparationsn}
\end{center}
\end{figure} 

\begin{figure}
\begin{center}
\begin{tabular}[b]{c}
    \epsfig{file=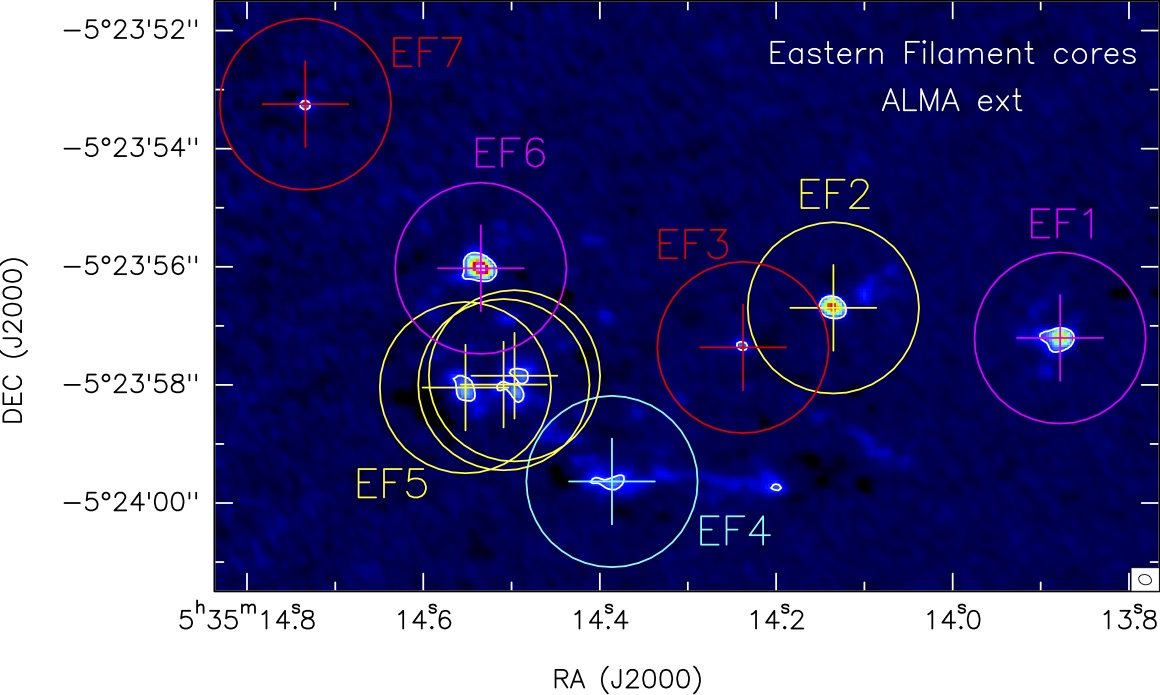, scale=0.4, angle=0}\\
\end{tabular}
\caption{Cores EF1 to EF7 identified in the Eastern Filament. Plus signs correspond to the positions of the ALMA sources (listed in Table~\ref{talma}), and the circles mark the distance of 556~AU with respect to each source.}
\label{fseparationsef}
\end{center}
\end{figure} 

\begin{figure}
\begin{center}
\begin{tabular}[b]{c}
    \epsfig{file=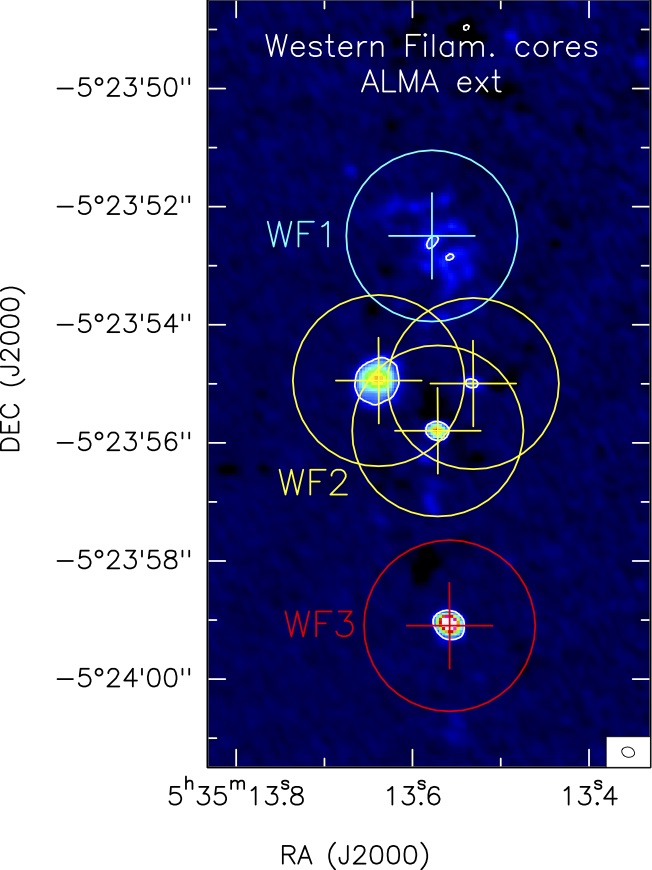, scale=0.3, angle=0}\\
\end{tabular}
\caption{Cores WF1, WF2 and WF3 identified in the Western Filament. Plus signs correspond to the positions of the ALMA sources (listed in Table~\ref{talma}), and the circles mark the distance of 556~AU with respect to each source.}
\label{fseparationswf}
\end{center}
\end{figure} 

\begin{figure}
\begin{center}
\begin{tabular}[b]{c}
    \epsfig{file=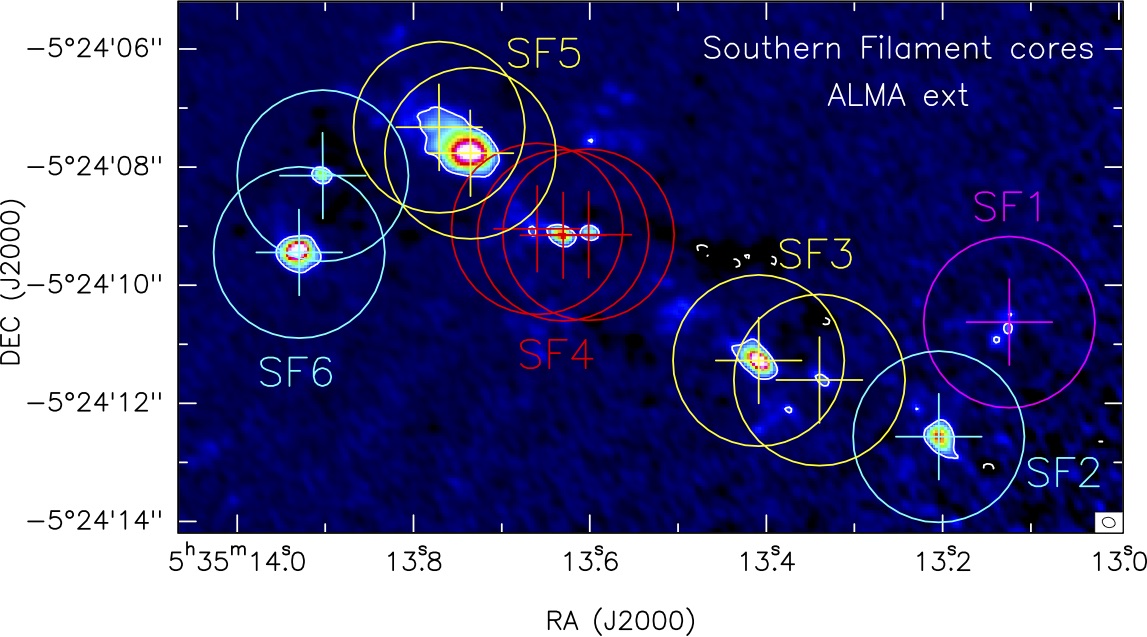, scale=0.4, angle=0}\\
\end{tabular}
\caption{Cores SF1 to SF6 identified in the Southern Filament. Plus signs correspond to the positions of the ALMA sources (listed in Table~\ref{talma}), and the circles mark the distance of 556~AU with respect to each source.}
\label{fseparationssf}
\end{center}
\end{figure}


\end{document}